\documentclass[10pt,twocolumn]{IEEEtran}
\usepackage{graphicx}
\usepackage{subfig}
\usepackage{cite}
\usepackage{amsmath,bm,amsfonts}
\usepackage{picins}
\usepackage{url,color}
\usepackage{epsfig,multirow,algorithm,algpseudocode,setspace,float}
\usepackage{verbatim}
\usepackage{lineno,hyperref}
\usepackage{booktabs}
\usepackage{xcolor}
\newcommand{\card}[1]{\color{#1}{\rule{2em}{0.7em}}}

\modulolinenumbers[5]
\newcommand{\tabincell}[2]{\begin{tabular}{@{}#1@{}}#2\end{tabular}}
\definecolor{myred}{RGB}{255,0,0}
\definecolor{mygreen}{RGB}{0,255,0}
\definecolor{myblue}{RGB}{0,0,255}
\definecolor{myyellow}{RGB}{255,255,0}
\definecolor{mypurple}{RGB}{255,0,255}
\definecolor{mycyan}{RGB}{0,255,255}
\definecolor{mybrown}{RGB}{200,100,0}
\definecolor{mygrassgreen}{RGB}{0,200,100}
\definecolor{mybluepurple}{RGB}{100,0,200}
\definecolor{mycrimson}{RGB}{200,0,100}
\definecolor{mylightgreen}{RGB}{100,200,0}
\definecolor{mydarkblue}{RGB}{0,100,200}
\definecolor{mydarkbrown}{RGB}{150,75,75}
\definecolor{myblackishgreen}{RGB}{75,150,75}
\definecolor{mydeeppurple}{RGB}{75,75,150}
\definecolor{mypink}{RGB}{255,100,100}

\begin{document}

\title{{LiteDepthwiseNet: An Extreme Lightweight Network for Hyperspectral Image Classification}}

\author{Benlei Cui, Xue-Mei Dong, Qiaoqiao Zhan, Jiangtao Peng, Weiwei Sun

\IEEEcompsocitemizethanks{\IEEEcompsocthanksitem
This work was supported in part by First Class Discipline of Zhejiang-A (Zhejiang Gongshang University - Statistics), National Natural Science Foundation of China under Grant Nos. 11701509, 61871177, 11771130, 41971296, by the Zhejiang Provincial Natural Science Foundation of China (LR19D010001), and by the Open Fund of State Laboratory of Information Engineering in Surveying, Mapping and Remote Sensing, Wuhan University under Grant 18R05. Corresponding author: Xue-Mei Dong
\IEEEcompsocthanksitem
B. Cui, X. Dong, and Q. Zhan are with the School of Statistics and Mathematics, Zhejiang Gongshang University, Hangzhou 310018, P. R. China
(email: dongxuemei@zjgsu.edu.cn).

J. Peng is with the Hubei Key Laboratory of Applied Mathematics, Faculty of Mathematics and Statistics,
Hubei University, Wuhan 430062, P. R. China. (e-mail: pengjt1982@hubu.edu.cn)

W. Sun is with the Department of Geography and Spatial Information Techniques,
Ningbo University, Ningbo 315211, P. R. China (e-mail: sunweiwei@nbu.edu.cn)

}}

\maketitle

\begin{abstract}
Deep learning methods have shown considerable potential for hyperspectral image (HSI) classification, which can achieve high accuracy compared with traditional methods. However, they often need a large number of training samples and have a lot of parameters and high computational overhead. To solve these problems, this paper proposes a new network architecture, LiteDepthwiseNet, for HSI classification. Based on 3D depthwise convolution, LiteDepthwiseNet can decompose standard convolution into depthwise convolution and pointwise convolution, which can achieve high classification performance with minimal parameters. Moreover, we remove the ReLU layer and Batch Normalization layer in the original 3D depthwise convolution, which significantly improves the overfitting phenomenon of the model on small sized datasets. In addition, focal loss is used as the loss function to improve the model's attention on difficult samples and unbalanced data, and its training performance is significantly better than that of cross-entropy loss or balanced cross-entropy loss. Experiment results on three benchmark hyperspectral datasets show that LiteDepthwiseNet achieves state-of-the-art performance with a very small number of parameters and low computational cost.
\end{abstract}


\IEEEpeerreviewmaketitle

\section{Introduction}

\IEEEPARstart{I}{n} recent years, as a kind of optical remote sensing image with high spectral resolution, hyperspectral images (HSIs) have attracted much attention because of their unique properties and the massive information they contain \cite{Li2019}. HSIs have been widely used in mineral exploration \cite{Yokoya2016}, soil salinity estimation \cite{Yang2017}, and anomaly detection \cite{Xie2019}. At present, HSI classification is one of the most active research areas in the remote sensing field. Its goal is to classify each marked pixel in the image correctly, where a pixel can be regarded as a high-dimensional vector with each entry corresponding to the spectral reflectance in a specific wavelength \cite{Li2019Overview}.

Early work on HSI classification mainly focused on exploring spectral feature information and ignoring spatial information \cite{Qian2001,Melgani2004,Li2010,Zhong2012,Peng2015}, which often leads to poor generalization performance. After this, some studies began to incorporate spatial contextual information into pixel-level classifiers. Representative work includes extended morphological profiles (EMP) \cite{Benediktsson2005,Li2013}, edge-preserving filtering \cite{Kang2014} and sparse representation models \cite{Chen2011,Fang2014,Fang2017,Peng2019,Peng2017}, etc. Due to the spatial homogeneous characteristics of HSI, spectral-spatial HSI classification methods usually show better results than spectral-based methods. Notwithstanding, most of spectral-spatial methods depend on hand-crafted or shallow features, which are usually designed for specific tasks and rely on expert knowledge. This largely limits the flexibility and applicability of these methods in difficult situations.

Recently, deep learning (DL) methods have been applied in HSI classification successfully. The earlier DL-based HSI classification methods based on fully connected neural networks, such as stacked autoencoders (SAEs) \cite{Chen2014} and recursive autoencoders (RAEs) \cite{Zhang2017}, can only handle one-dimensional vectors. Thus, the spatial structure information of an HSI is destroyed. The emergence of convolutional neural networks (CNNs) overcomes this defect. Several 2D CNN based architectures were proposed, including R-VCANet \cite{Pan2017} and bayesian 2D CNN \cite{Cao2018}, etc. However, a HSI has too many channels, which often causes 2D convolution kernels to be too deep, and there is also a significant increase in the number of parameters. Therefore, 3D CNN based HSI classification methods were proposed. Lee et al. \cite{lee2016} described a contextual deep CNN by applying multiple 3D local convolutional filters with different sizes to jointly exploit the spatial and spectral features of a HSI. Chen et al. \cite{chen2016} established a 3D CNN-based feature extraction model with regularization to extract the effective spectral-spatial features of HSIs. Hamida et al. \cite{Benhamida2018} proposed a new 3D deep learning method that can simultaneously process spectral and spatial information while incurring lower computation cost (i.e. floating point operations, FLOPs). Although 3D CNN can reduce the number of parameters, it still incurs more calculational consumption because it needs to traverse in depth and lacks a global view of spectral information. HybridSN, a recent work proposed by Roy et al. in \cite{Roy2020}, combined 2D CNN and 3D CNN, where 3D CNN was used to mine spatial and spectral information and 2D CNN integrated the information to output the prediction result. It can effectively improve classification accuracy and has a global view to integrate spectral features. However, without an end-to-end architecture, PCA was still used to compress the redundant spectral information, which adds computational cost.

Although the aforementioned DL models have shown excellent HSI classification performance, they usually need a large number of training samples and network parameters, and also have higher computational cost. For HSI classification, the available labeled pixels are usually very limited due to the difficulty in collecting and cost in labeling. Moreover, the distribution of categories in these labeled data is imbalanced. Constructing robust DL models with high performance and low computational cost based on imbalanced and small sample-sized data is a huge challenge in this field.
Recently, a lightweight neural network, LiteDenseNet \cite{Li2020}, has been proposed for HSI classification.
Different from traditional 3D CNN architectures, LiteDenseNet has a 3D two-way dense structure and uses group convolution,
which greatly reduces the number of parameters and computational cost.
However, the group convolution cuts off the connection of different channels and may lead to a loss of accuracy.
To solve this problem and further reduce the parameters,
we propose an extremely lightweight network, LiteDepthwiseNet, for HSI classification tasks. Its main advantages are summarized as follows.

\begin{enumerate}
\item A 3D depthwise convolution is used to replace the group convolution. The pointwise convolution in the 3D depthwise convolution can connect all hyperspectral channels, and the corresponding network architecture has a full-channel receptive field and is more suitable for HSI classification task. Based on this new network architecture, LiteDepthwiseNet only involves a very small number of parameters and FLOPs.
\item Focal loss (FL) is introduced to replace mainstream cross entropy loss (CEL) as the loss function. It increases the attention on small sample categories and difficult-to-classify samples, which helps to improve the ultimate performance of the model.
\item The middle activation layer and normalization layer in the original 3D depthwise convolution network are stripped, which can enhance the linearity of the model, so as to reduce the overfitting phenomenon when there are few training samples for HSIs.
\end{enumerate}

The rest of this paper is arranged as follows. In Section \ref{II}, we briefly introduce the related work. The structure of the proposed LiteDepthwiseNet is detailed in Section \ref{III}. The comparison experiment results of seven algorithms on three well-known HSI datasets are reported in Section \ref{IV}. Section \ref{V} summarizes this paper.

\section{Related work}\label{II}
This section describes some related concepts and gives a brief introduction to LiteDenseNet, which inspires the main design of the proposed network architecture in this paper.

\subsection{2D depthwise separable convolution}
\begin{figure}[H]
	\centering
	\includegraphics[width=1\linewidth,height=0.32\linewidth]{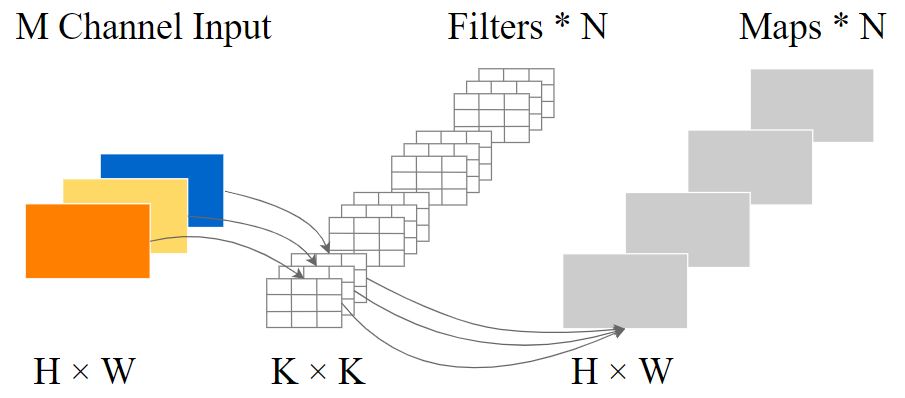}
	\caption{A standard convolution.}
	\label{figure1}
\end{figure}

\begin{figure}[H]
	\centering
	\includegraphics[width=1\linewidth,height=0.3\linewidth]{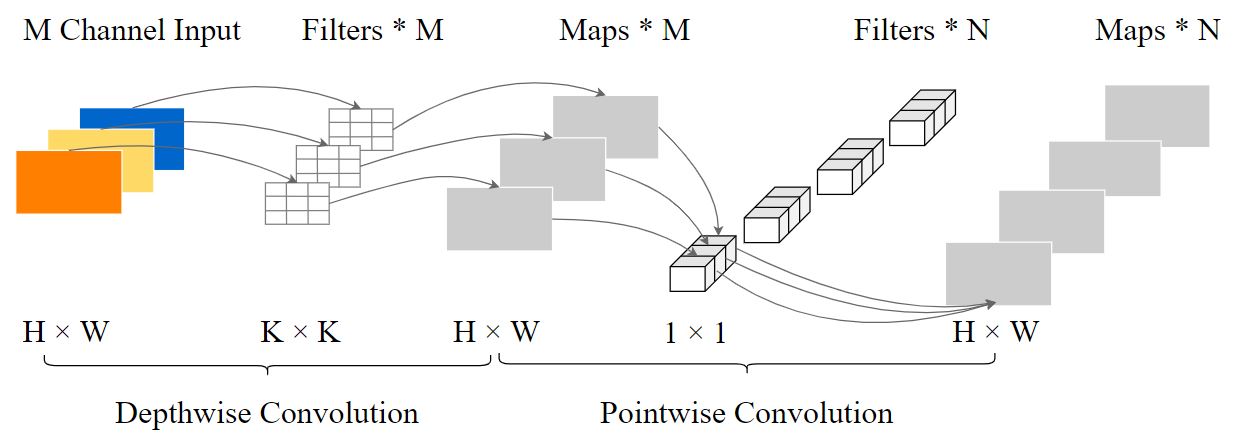}
	\caption{A depthwise separable convolution.}
	\label{figure2}
\end{figure}

Different from standard convolution, the depthwise separable convolution proposed in \cite{Howard2017MobileNets} includes a depthwise convolution, i.e. a spatial convolution performed independently over each channel of an input, followed by a pointwise convolution, i.e. a $ 1 \times 1 $ convolution, projecting the channels output by the depthwise convolution onto a new channel space. Figs. \ref{figure1} and Figs. \ref{figure2} show a simple schematic of a standard convolution and a depthwise separable convolution, respectively.

By splitting the standard convolution into two processes of depthwise convolution and pointwise convolution, the computational cost and the number of parameters in the 2D convolution can be significantly reduced under certain conditions.

\subsection{Cross entropy loss}
The CEL function is a loss function often used in classification problems. It is an important concept in information theory, where it is mainly used to measure the difference between two probability distributions.

The standard CEL for classification problems is,
\begin{equation}\label{formula1}
\text{CEL}(p,y)=
\begin{cases}
- \text{log}(p) & \text{if} ~ y=1 \\
- \text{log}(1-p) & \text{otherwise}.
\end{cases}
\end{equation}
where $ p $ represents the predicted probability of a sample in the category, and $ y $, an indicate variable equal to 0 or 1, represents whether the sample is correctly classified. It can be observed that when $y$ is $ 1 $, the closer the $ p $ is to $ 1 $, the smaller the loss and if $y$ is $ 0 $, the closer the $ p $ is to 0, the smaller the loss, which is in line with the direction of optimization.

To simplify the expression, we introduce a piecewise function $ F $ as:
\begin{equation}\label{formula2}
F_y(t)=
\begin{cases}
t & \text{if} ~ y=1 \\
1-t & \text{otherwise}.
\end{cases}
\end{equation}
and then the standard CEL can be reformulated as:
\begin{equation}\label{formula3}
\text{CEL}(p,y) = - \text{log}( F_y(p) )
\end{equation}

In order to handle data with imbalanced categories, a coefficient $ \alpha\in [0,1] $ is added to the standard cross-entropy loss function to balance the weights of samples of different categories, which results in the following balanced CEL (BCEL):
\begin{equation}\label{formula4}
\text{BCEL}(p,\alpha, y)=-F_y(\alpha) \text{log}(F_y(p))
\end{equation}
\subsection{LiteDenseNet}

\begin{figure}[H]
  \centering
  \includegraphics[scale=0.4]{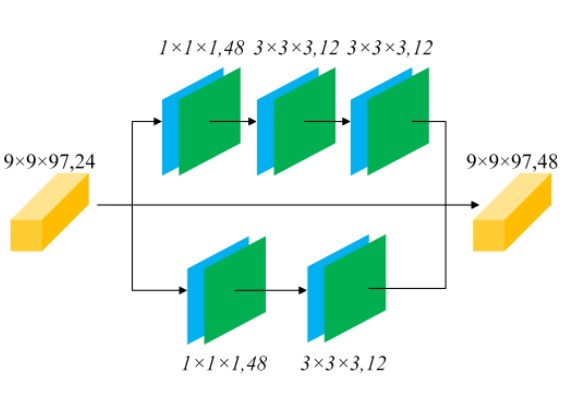}
  \caption{The 3D two-way dense layer in \cite{Li2020}.}
  \label{figure3}
\end{figure}

In 2020, Li et al. \cite{Li2020} proposed a lightweight network LiteDenseNet. To obtain receptive fields in different scales, the network uses a 3D two-way dense layer which was proposed in PeleeNet \cite{Pelee2018} and DBDA \cite{Li2020Classification}. Its schematic diagram is shown in Fig. \ref{figure3}.

This 3D two-way dense layer comprises a top layer and a bottom layer. After performing a $ 1 \times 1 \times 1 $ convolution transformation on the input data, the top layer uses a two stacked $ 3 \times 3 \times 3 $ convolution to capture global features and obtains an output with size $ (9 \times 9 \times 97, 12) $,  and the bottom layer uses a convolution kernel with size $ 3 \times 3 \times 3 $ to capture local features and obtains an output with size $ (9 \times 9 \times 97, 12) $. The outputs of the two-way dense layer are used to perform the concatenate operation to generate a 3D-cube with size $ (9 \times 9 \times 97, 48) $.

\begin{figure}[H]
  \centering
  \includegraphics[scale=0.14]{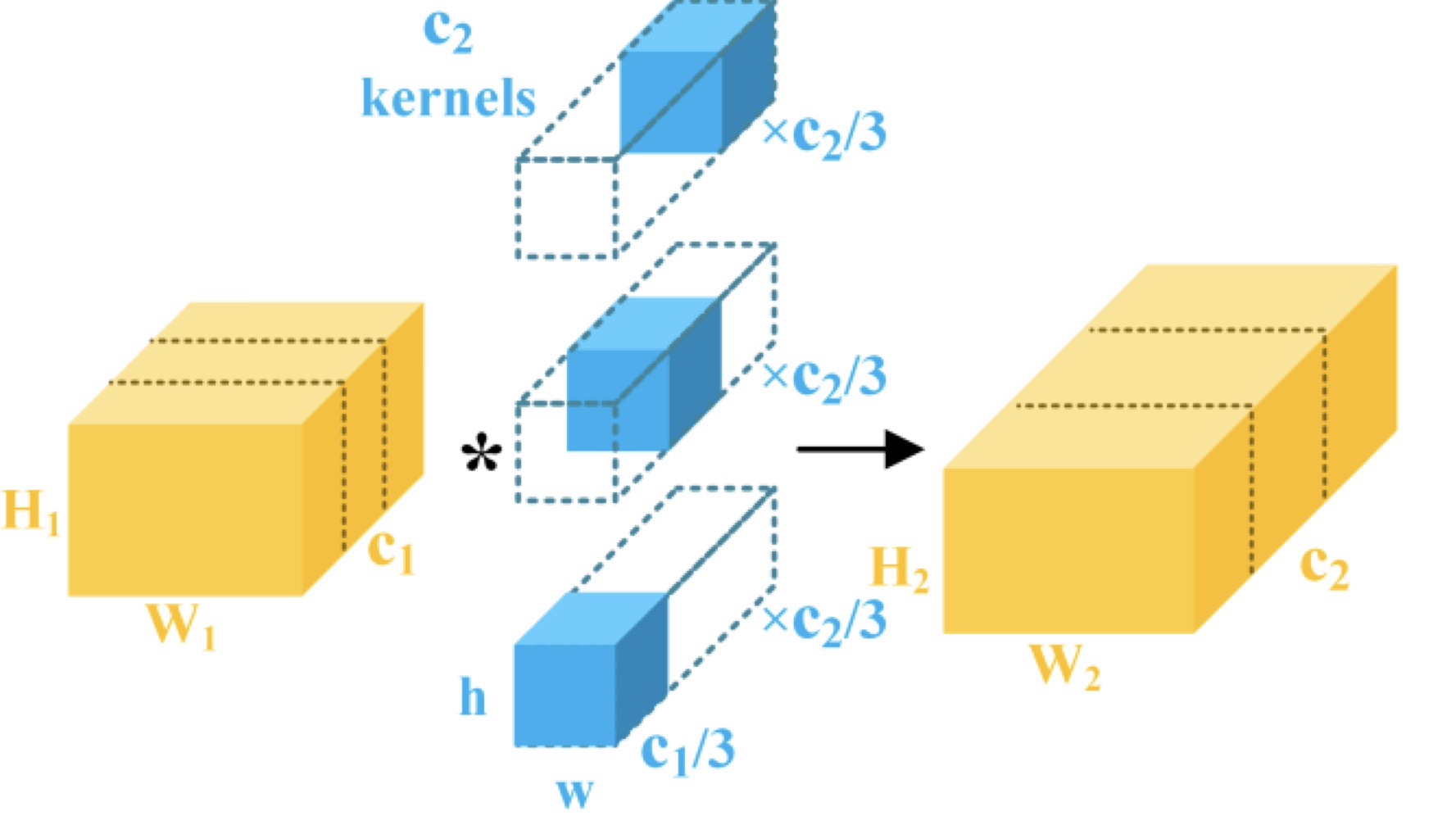}
  \caption{Group Convolution.}
  \label{figure4}
\end{figure}

However, the 3D convolutional block in the original 3D two-way dense layer is a computationally intensive model. In order to reduce the number of parameters and computational cost, LiteDenseNet uses group convolution instead of normal convolution. The group convolution given in \cite{Li2020} is shown in Fig. \ref{figure4}.

Given an input data with size $ h \times w \times l$ and $ c_F $ channels, the normal convolution layer tends to have the same channel size as the inputs, whereas the group convolution divides the original convolution kernels and the input into 3 groups by channel dimension. As a result, the input is divided into three groups with the size of  $ h \times w \times l $ and channels $c_F/3$. Three groups are fed into a normal convolution with the $c_F/3$ channels separately. So the channel size is reduced from $ h \times w \times l \times c_F $ to $ h \times w \times l \times (c_F/3)$. In this way, group convolution significantly reduces the number of parameters and computational cost while maintaining the size of the input and the output.

\section{The proposed LiteDepthwiseNet}\label{III}
In this section, we describe the proposed LiteDepthwiseNet in detail.
\subsection{Modified 3D Depthwise Convonlution}

\begin{figure}[H]
  \centering
  \includegraphics[scale=0.4]{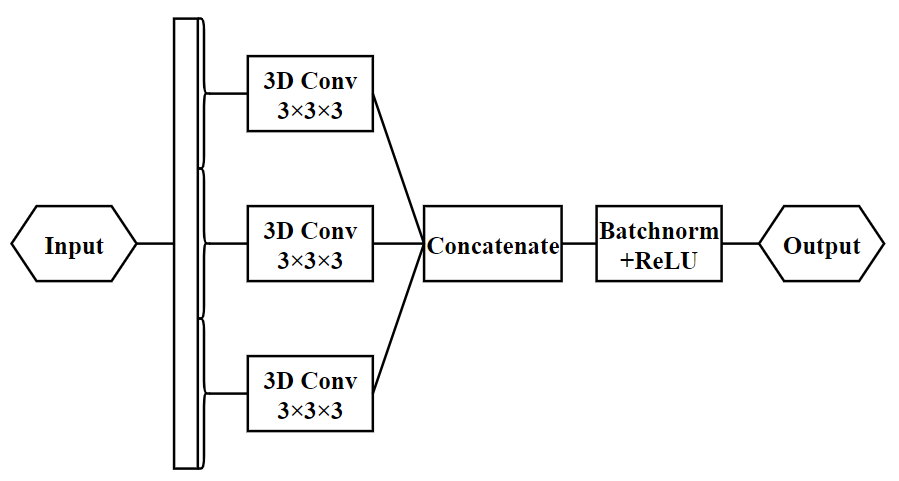}
  \caption{Group convolution with 3 groups.}
  \label{figure5}
\end{figure}

\begin{figure}[H]
  \centering
  \includegraphics[scale=0.4]{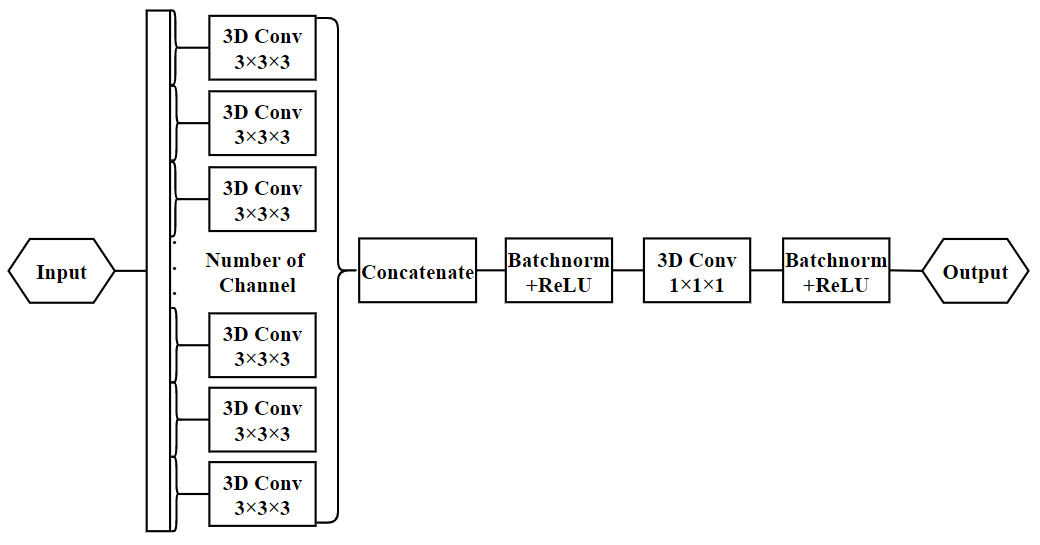}
  \caption{3D depthwise convolution.}
  \label{figure6}
\end{figure}

\begin{figure}[H]
  \centering
  \includegraphics[scale=0.4]{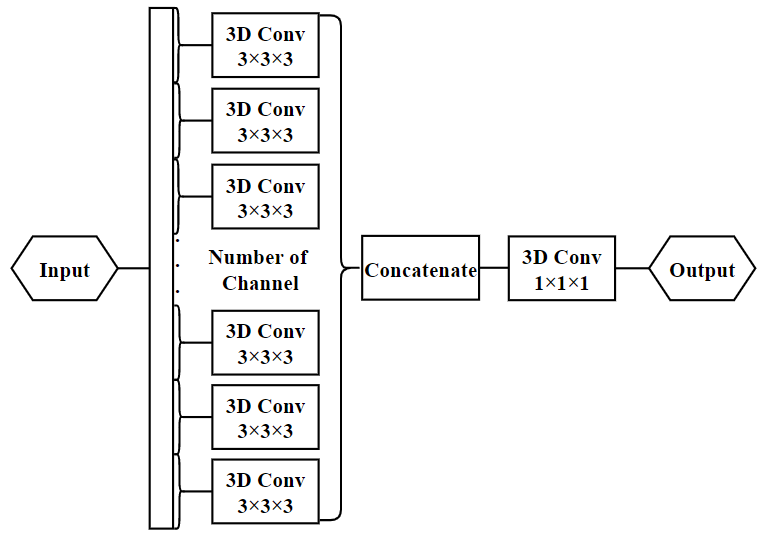}
  \caption{Modified 3D depthwise convolution.}
  \label{figure7}
\end{figure}

LiteDenseNet reduces the parameters and computational cost using group convolution to replace standard convolution. However, group convolution permanently cuts off some connections between channels, which results in reducing the classification performance to some extent. Moreover, for HSI classification, there is still room to reduce the parameters and computational cost. So, we consider replacing group convolution with depthwise separable convolution, in which pointwise convolution can connect all the channels. At the same time, for HSIs, there are too many channels, so 2D depthwise separable convolution will still have too many parameters. A good choice is to use 3D depthwise convolution originally proposed in \cite{Ye20193D} for RGB images. At this point, a new problem appears. The limited number of training samples for HSIs always causes overfitting for the model which has a too strong nonlinear ability. So, we propose a modified 3D depthwise convolution for HSI classification. The major adjustment is that the middle activation layer and normalization layer of the original 3D depthwise convolution network are stripped, which enhances the linearity of the model and reduces the overfitting phenomenon. Moreover, the number of parameters and the computational consumption are greatly reduced. Structural comparison diagrams of group convolution, 3D depthwise convolution and modified 3D depthwise convolution are shown in Figs. \ref{figure5}, \ref{figure6}, and \ref{figure7}, respectively.

\subsection{Focal Loss}

Since most HSIs have imbalanced data categories \cite{Li2019Overview}, using the standard CEL function for training will lead to low classification accuracy for the small sample categories. The BCEL function is commonly used to deal with this situation. However, the BCEL function is also likely to focus on the easy-to-classify samples. In order to improve the classification performance, the model should focus on both small-category samples and difficult-to-classify samples. For this purpose, a focal loss is proposed in \cite{Lin2017Focal} and expressed as
\begin{equation}\label{formula5}
{\rm FL}(p, \alpha, \gamma, y) = - F_y(\alpha) (1 - F_y(p))^\gamma \log(F_y(p)),
\end{equation}
where $\alpha\in [0,1], \gamma\geq 0.$

From formula (\ref{formula5}), it is easy to know the following three points. Firstly, similar to the BCEL, parameter $\alpha $ is introduced to reflect the imbalanced problem of different categories of samples, which helps to improve the accuracy of the final model. Secondly, the modulating factor $ (1 - F_y(p))^\gamma $ is added to adjust the weights of easy-to-classify samples and difficult-to-classify samples. When a sample is misclassified and $ F_y(p) $ is small, then $ (1 - F_y(p))^\gamma $ is close to 1, and the loss is not affected. However, if $ F_y(p) \to 1 $, which indicates that the classification prediction of the sample is better, $ (1 - F_y(p))^\gamma $ is 0, and the loss is down-weighted. Lastly, the focusing parameter $ \gamma $ smoothly adjusts the down-weighted rate of easy-to-classify samples. When $\gamma = 0 $, the FL is equivalent to the BCEL and with the increase of $ \gamma $, the effect of the modulating factor also increases. Of these, $ \alpha $ can be set by category frequency, or it can be obtained as a hyperparameter for cross-validation.
\subsection{The Framework of the LiteDepthwiseNet}

\begin{figure}[p]
  \centering
  \includegraphics[scale=0.3]{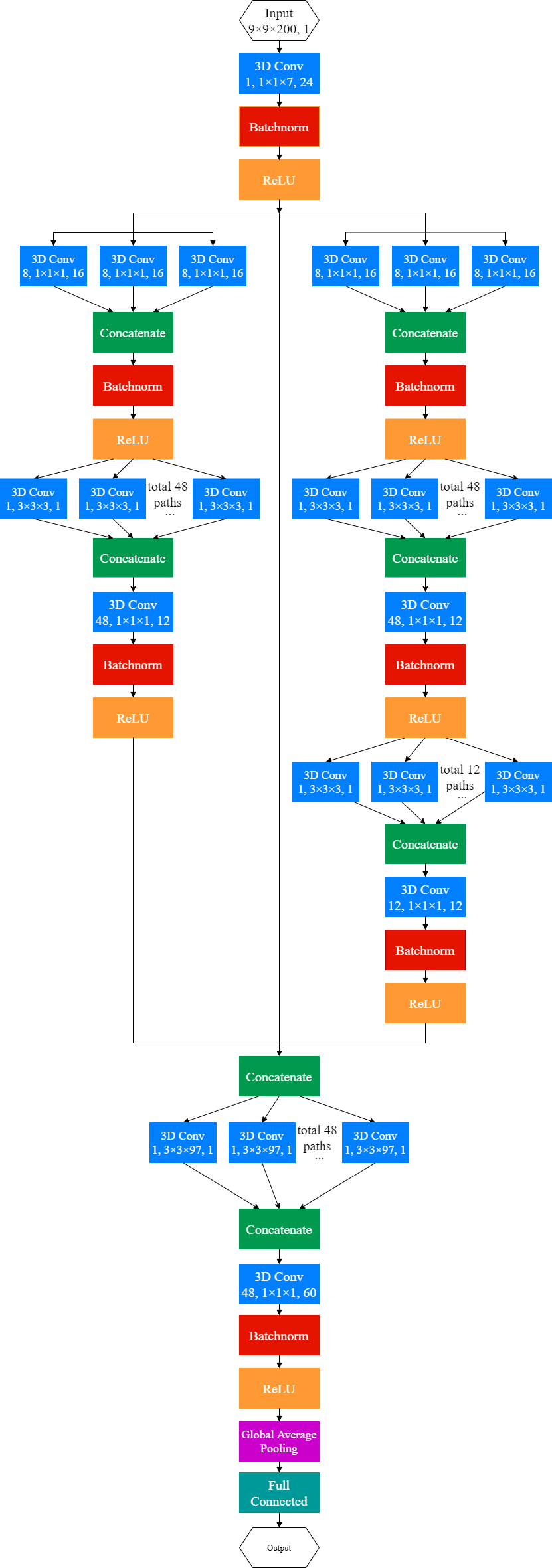}
  \caption{The LiteDepthwiseNet architecture}
  \label{figure8}
\end{figure}

The overall framework of the proposed LiteDepthwiseNet and the specific parameters in an example are shown in Fig. \ref{figure8} and Table \ref{table1}, respectively.

Specifically, in our network framework, batch normalization (BN) and ReLU activation (ReLU) are added after each point convolution and ordinary convolution. For the sake of brevity, it will not be specified in the following description. When an input, a 3D-cube of size $ (9 \times 9 \times 200,1) $, is fed into a 3D-CNN layer with size $ (1 \times 1 \times 7,24) $, we get an output with size of $ (9 \times 9 \times 97,24) $. Next we put the output into two branches respectively. The right branch (Branch 1) performs group convolution on the input and divides the input into three groups, where each group is a 3D-cube with size of $ (9 \times 9 \times 97,8) $. Then, we put them into a 3D-CNN layer with size of $ (1 \times 1 \times 1, 16) $. Until now, we have retained the original design of LiteDenseNet, because modifying the first and second layers to 3D depthwise convolution increases the number of parameters. After this, the output results of the three 3D-CNNs are stacked by channel dimension, and we obtain the output with size of $ (9 \times 9 \times 97,48) $. The subsequent modules are all composed of 3D depthwise convolution, which divides $ (9 \times 9 \times 97,48) $ into 48 independent 3D blocks by channel dimension, and uses 48 3D-CNN filters with the size of $ (3 \times 3 \times 3,1) $ for one-to-one convolution operation, and the size of output is $ (9 \times 9 \times 97,48) $. We feed this into the 3D-CNN with the size of $ (1 \times 1 \times 1,12) $ for a point-by-point linear combination, which makes up for the defect of group convolution. Then, we continue to feed a 3D depthwise convolution, where the number of input and output channels are both 12. At this point, the design of the right branch is completed. The left branch (Branch 2) feeds the input obtained by the group convolution into a 3D depthwise convolution. Finally, we contact the outputs of the three channels, that is, the input that has not entered any branch with the size of $ (9 \times 9 \times 97,24) $, the output of the right branch with the size of $ (9 \times 9 \times 97,12) $, and the output of the left branch with the size of $ (9 \times 9 \times 97,12) $, to obtain the output with the size of $ (9 \times 9 \times 97,48) $, which is fed into the last 3D depthwise convolution, and the final result is obtained through a global average pooling layer and a fully connected layer.

\begin{table}[!ht]
\caption{THE IMPLEMENTATION DETAILS ON AN EXAMPLE OF LITEDEPTHWISENET}
\scalebox{0.75}{
\begin{tabular}{ccccc}
  \toprule
  \toprule
  \multicolumn{2}{c}{Layer name} & Kernel Size & Group & Output Size \\
  \hline
  \multicolumn{2}{c}{Input} & - & - & $(9\times9\times200,1)$ \\
  \hline
  \multicolumn{2}{c}{3D-CNN+BN+ReLU} & $(1\times1\times7)$ & 1 & $(9\times9\times97,24)$ \\
  \hline
  \multirow{3}*{Branch 1} & \tabincell{c}{3D-CNN+Concatenate\\+BN+ReLU} & $(1\times1\times1)$ & 3 & $(9\times9\times97,48)$ \\
  \cline{2-5}
  & 3D-CNN+Concatenate & $(3\times3\times3)$ & 48 & $(9\times9\times97,48)$ \\
  \cline{2-5}
  & 3D-CNN+BN+ReLU & $(1\times1\times1)$ & 1 & $(9\times9\times97,12)$ \\
  \hline
  \multirow{5}*{Branch 2} & \tabincell{c}{3D-CNN+Concatenate\\+BN+ReLU} & $(1\times1\times1)$ & 3 & $(9\times9\times97,48)$ \\
  \cline{2-5}
  & 3D-CNN+Concatenate & $(3\times3\times3)$ & 48 & $(9\times9\times97,48)$ \\
  \cline{2-5}
  & 3D-CNN+BN+ReLU & $(1\times1\times1)$ & 1 & $(9\times9\times97,12)$ \\
  \cline{2-5}
  & 3D-CNN+Concatenate & $(3\times3\times3)$ & 12 & $(9\times9\times97,12)$ \\
  \cline{2-5}
  & 3D-CNN+BN+ReLU & $(1\times1\times1)$ & 1 & $(9\times9\times97,12)$ \\
  \hline
  \multicolumn{2}{c}{Concatenate} & - & - & $(9\times9\times97,48)$ \\
  \hline
  \multicolumn{2}{c}{3D-CNN+Concatenate} & $(3\times3\times97)$ & 48 & $(9\times9\times1,48)$ \\
  \hline
  \multicolumn{2}{c}{3D-CNN+BN+ReLU} & $(1\times1\times1)$ & 1 & $(9\times9\times1,60)$ \\
  \hline
  \multicolumn{2}{c}{Global Average Pooling} & - & - & $(1\times60)$ \\
  \hline
  \multicolumn{2}{c}{Full Connected} & - & - & $(1\times16)$ \\
  \bottomrule
  \bottomrule
\end{tabular}}
\label{table1}
\end{table}

\subsection{Comparing Calculation Cost and Number of Parameters}

Under certain conditions, compared with group convolution, our modified 3D depthwise convolution reduces more parameters and has lower computation cost. For example, taking $ (h \times w \times l) , c_F$ as the size and the number of channels of the input, $ (h_F \times w_F \times l_F) $ as the convolution kernel size, $c_G$ as the size and the number of channels of the output, the number of parameters is given by
\begin{equation*}
(h \times w \times l \times (c_F/3)) \times (c_G/3) \times 3.
\end{equation*}
and the amount of the computation of group convolution is calculated as
\begin{equation*}
(h \times w \times l \times (c_F/3) \times h_F \times w_F \times l_F \times (c_G/3)) \times 3.
\end{equation*}

Whereas the number of parameters of the 3D depthwise convolution is as follows,
\begin{equation*}
h \times w \times l \times c_F \times 1 + 1 \times 1 \times 1 \times c_F \times c_G.
\end{equation*}
and the amount of the computation of 3D depthwise convolution is
\begin{equation*}
h \cdot w \cdot l \cdot c_F \cdot h_F \cdot w_F \cdot l_F + c_F \cdot c_G \cdot h_F \cdot w_F \cdot l_F.
\end{equation*}

For some special cases, such as $ 1\times1\times1 $ convolution kernel, the 3D depthwise convolution increases the number of parameters and the computation cost.

\begin{figure}[!t]
  \centering
  \includegraphics[scale=0.3]{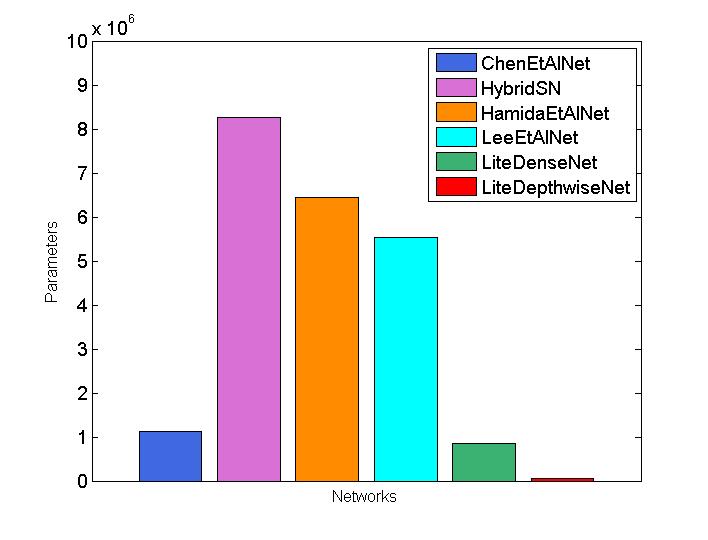}
  \caption{Parameters of each algorithm}
  \label{figure9}
\end{figure}

\begin{figure}[!t]
  \centering
  \includegraphics[scale=0.3]{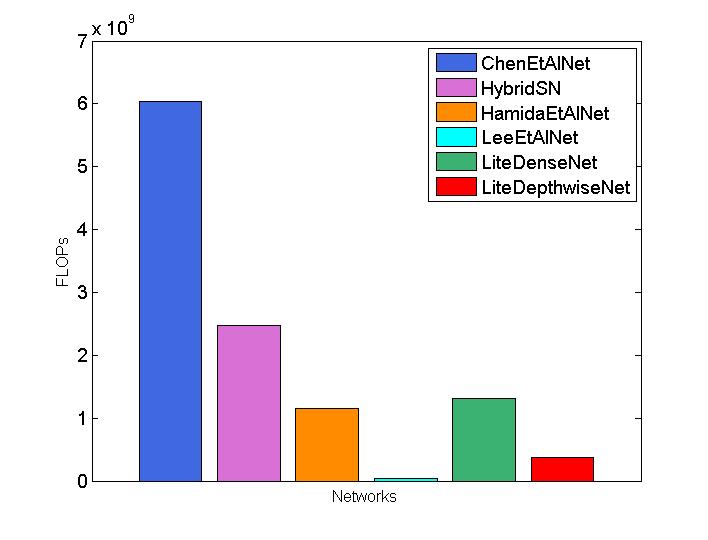}
  \caption{FLOPs of each algorithm}
  \label{figure10}
\end{figure}

Table \ref{table2}, Fig. \ref{figure9} and Fig. \ref{figure10} amount compare the results of the number of parameters and of the computation for our proposed LiteDepthwiseNet and five other deep learning algorithms when the input size is $ 25 \times 25 \times 200 $. Of these, for the HybridSN method, PCA is required for preprocessing. For the convenience of comparison, this operation is not used to calculate the computation overhead.

\begin{table}[H]
\centering
\caption{THE COMPARISON OF SIX ALGORITHMS}
\scalebox{0.9}{
\begin{tabular}{c|c|c|c}
  \toprule
  \toprule
  Network & Input size & FLOPs & Parameters \\
  \hline
  ChenEtAlNet & $ 25 \times 25 \times 200 $ & 6.030G & 1.120M \\
  HybridSN & $ 25 \times 25 \times 200 $ & 2.483G & 8.256M \\
  HamidaEtAlNet & $ 25 \times 25 \times 200 $ & 1.148G & 6.452M \\
  LeeEtAlNet & $ 25 \times 25 \times 200 $ & 37.819M & 5.532M \\
  LiteDenseNet & $ 25 \times 25 \times 200 $ & 1.316G & 852.309k \\
  LiteDepthwiseNet & $ 25 \times 25 \times 200 $ & {\bf 369.331M} & {\bf 51.616k} \\
  \bottomrule
  \bottomrule
\end{tabular}}
\label{table2}
\end{table}

We can see that the number of parameters of the proposed LiteDepthwiseNet is much smaller than that of the other five algorithms. In particular, it is 93.94\% less than the LiteDenseNet. Moreover, the computation cost of LiteDepthwiseNet is 71.91\% less than that of the LiteDenseNet. Note that the LeeEtAlNet has the minimal computation cost, but later experiment results show that its classification accuracy is far lower than ours.

\section{EXPERIMENT RESULTS}\label{IV}

In this section, we perform experiments on three widely used hyperspectral datasets to compare the performance of the proposed LiteDepthwiseNet with six other methods, namely, LiteDenseNet \cite{Li2020}, LeeEtAlNet \cite{lee2016}, HamidaEtAlNet \cite{Benhamida2018}, HybridSN \cite{Roy2020}, ChenEtAlNet \cite{chen2016} and the classic SVM. Except for LiteDenseNet, the other algorithms are tested using their publicly recognized open source codes or the codes provided by the authors. Since LiteDenseNet encapsulates its code, we use our own reproduced code with balanced cross-entropy loss for the experiments. All the  experiments are implemented on Ubuntu 18.04.3 LTS, CPU: i9 9900k, GPU: 2x2080ti, RAM: 64 GB, GPU memory: 22GB, Python: 3.6.1, Torch: 1.3.1. Three quantitative indicators, overall accuracy (OA), average accuracy (AA) and $\kappa$ coefficient (Kappa), are used to evaluate the performance of each method.

\subsection{Data Description}
The following three open HSI datasets are used for the experiments in this paper. They can be obtained from \url{http://www.ehu.eus/ccwintco/index.php?title=Hyperspectral_Remote_Sensing_Scenes}.

\textbf{Indian Pines (IP):} This dataset was acquired by the Airborne Visible Infrared Imaging Spectrometer (AVIRIS) sensor in north-western Indiana. The size of the scene is $ 145 \times 145 \times 200 $ with a spatial resolution 20 m per pixel and spectral coverage ranging from 0.4 to 2.5$ \mu m $. The ground truth (GT) available is divided into 16 classes of vegetation.

\textbf{University of Pavia (UP):} This dataset captures an urban area surrounding the University of Pavia, Italy, which was collected by the Reflective Optics Imaging Spectrometer (ROSIS-03) sensor in Northern Italy in 2001. The size of the scene is $ 610 \times 340 \times 103 $ with a spatial
resolution of 1.3 m per pixel and spectral coverage ranging from 0.43 to 0.86$ \mu m $. The ground truth is designated into 9 urban land-cover classes.

\textbf{Pavia Center (PC):} This dataset is also acquired by the Reflective Optics Imaging Spectrometer (ROSIS-03) sensor. The size of the scene is $ 1096 \times 715 \times 102 $ with a spatial resolution 1.3 m per pixel and spectral coverage ranging from 0.43 to 0.86$ \mu m $. The ground truth available is divided into 9 classes.

In the follow-up experiments, the \textbf{UP} and \textbf{PC} datasets are divided into training, verification and testing sets, and the ratio of training, verification and testing within each class is the same. Here, we choose 0.5\% of the \textbf{UP} data for training and 0.1\% of the \textbf{PC} data for training, and the number of samples in the training set and verification set of the \textbf{UP} and \textbf{PC} datasets are the same. But for the \textbf{IP} data, the training ratio is too low, which will leads to the absence of training samples in some categories. To ensure that there are at least 5 samples in each category, we use 5\% of the \textbf{IP} data for training and do not set a verification set.The detailed division information of the three datasets are listed in Table \ref{table3}, Table \ref{table4} and Table \ref{table5}, respectively.

\begin{table}[H]
\centering
\caption{THE SAMPLES FOR EACH CATEGORY OF TRAINING, VALIDATION AND TESTING FOR THE \textbf{IP} DATASET.}
\scalebox{0.74}{
\begin{tabular}{c|c|c|c|c|c}
  \toprule
  \toprule
  No. & Class & Total number & Training & Validation & Testing \\
  \hline
  1 & Alfal & 46 & 5 & - & 41 \\
  2 & Corn-notill & 1428 & 71 & - & 1357 \\
  3 & Corn-mintill & 830 & 41 & - & 789 \\
  4 & Corn & 237 & 12 & - & 225 \\
  5 & Grass-pasture &483 & 24 & - & 459 \\
  6 & Grass-trees & 730 & 37 & - & 693 \\
  7 & Grass-pasture-mowed & 28 & 5 & - & 23 \\
  8 & Hay-windrowed & 478 & 24 & - & 454 \\
  9 & Oats & 20 & 5 & - & 15 \\
  10 & Soybean-notill & 972 & 49 & - & 923 \\
  11 & Soybean-mintill & 2455 & 109 & - & 2346 \\
  12 & Soybean-clean & 593 & 30 & - & 563 \\
  13 & Wheat & 205 & 12 & - & 193 \\
  14 & Woods &1265 & 63 & - & 1202 \\
  15 & Buildings-Grass-Trees &386 & 19 & - & 367 \\
  16 & Stone-Steel-Towers & 93 & 6 & - & 87 \\
  \hline
     & Total & 10249 & 512 & - & 9737 \\
  \bottomrule
  \bottomrule
\end{tabular}}
\label{table3}
\end{table}

\begin{table}[H]
\centering
\caption{THE SAMPLES FOR EACH CATEGORY OF TRAINING, VALIDATION AND TESTING FOR THE \textbf{UP} DATASET.}
\scalebox{0.75}{
\begin{tabular}{c|c|c|c|c|c}
  \toprule
  \toprule
  No. & Class & Total number & Training & Validation & Testing \\
  \hline
  1 & Asphalt & 6631 & 33 & 33 & 6565 \\
  2 & Meadows & 18649 & 93 & 93 & 18463 \\
  3 & Gravel & 2099 & 10 & 11 & 2078 \\
  4 & Trees & 3064 & 15 & 16 & 3033 \\
  5 & Painted metal sheets & 1345 & 7 & 6 & 1332 \\
  6 & Bare Soil & 5029 & 25 & 25 & 4979 \\
  7 & Bitumen & 1330 & 7 & 6 & 1317 \\
  8 & Self-Blocking Bricks & 3682 & 19 & 19 & 3644 \\
  9 & Shadows  & 947 & 5 & 5 & 937 \\
  \hline
     & Total & 42776 & 214 & 214 & 42348 \\
  \bottomrule
  \bottomrule
\end{tabular}}
\label{table4}
\end{table}

\begin{table}[H]
\centering
\caption{THE SAMPLES FOR EACH CATEGORY OF TRAINING, VALIDATION AND TESTING FOR THE \textbf{PC} DATASET.}
\scalebox{0.75}{
\begin{tabular}{c|c|c|c|c|c}
  \toprule
  \toprule
  No. & Class & Total number & Training & Validation & Testing \\
  \hline
  1 & Water & 65971 & 65 & 65 & 65841 \\
  2 & Trees & 7598 & 7 & 7 & 7584 \\
  3 & Asphalt & 3090 & 3 & 3 & 3084 \\
  4 & Self-Blocking Bricks & 2685 & 3 & 3 & 2679 \\
  5 & Bitumen & 6584 & 6 & 6 & 6572 \\
  6 & Tiles & 9248 & 9 & 9 & 9230\\
  7 & Shadows & 7287 & 7 & 7 & 7273 \\
  8 & Meadows & 42826 & 42 & 42 & 42742 \\
  9 & Bare Soil & 2863 & 3 & 3 & 2857 \\
  \hline
     & Total & 148152 & 145 & 145 & 147862 \\
  \bottomrule
  \bottomrule
\end{tabular}}
\label{table5}
\end{table}

\subsection{Classification Maps and Categorized Results}

\subsubsection{Classification Maps and Categorized Results for the \textbf{IP} Dataset}

\begin{figure}[!t]
\centering
\subfloat[False color image]{
\includegraphics[width=2.5cm]{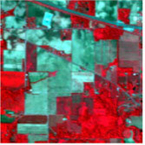}
}
\hfil
\subfloat[GT]{
\includegraphics[width=2.5cm]{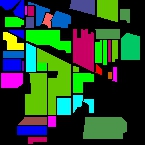}
}
\hfil
\subfloat[SVM]{
\includegraphics[width=2.5cm]{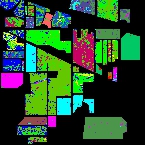}
}
\hfil
\subfloat[ChenEtAlNet]{
\includegraphics[width=2.5cm]{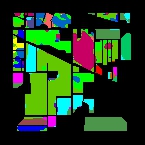}
}
\hfil
\subfloat[HybridSN]{
\includegraphics[width=2.5cm]{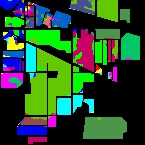}
}
\subfloat[HamidaEtAlNet]{
\includegraphics[width=2.5cm]{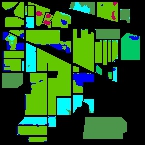}
}
\hfil
\subfloat[LeeEtAlNet]{
\includegraphics[width=2.5cm]{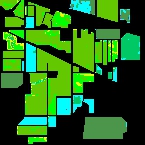}
}
\hfil
\subfloat[LiteDenseNet]{
\includegraphics[width=2.5cm]{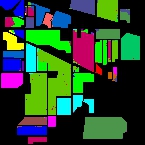}
}
\hfil
\subfloat[LiteDepthwiseNet]{
\includegraphics[width=2.5cm]{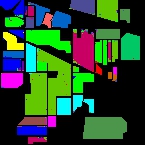}
}
\centering
\caption{Classification maps for the \textbf{IP} dataset using $ 5 \% $ of training samples. (a) False-color image. (b) GT. (c)-(i) The classification maps with disparate
algorithms.}
\label{figure11}
\end{figure}

The results of using different algorithms to classify the \textbf{IP} dataset are shown in Table \ref{table6}, and the classification maps of the different algorithms and the GT categories are shown in Fig. \ref{figure11}. The main characteristic of IP data is that the amount of data is small and the data distribution is imbalanced. In particular, the amount of data in categories 1, 7, 9 and 16 is less than 100, far less than that in other categories. Using the training data in Table \ref{table3}, the proposed algorithm performs best in OA and Kappa compared with other six algorithms. On most categories, LiteDepthwiseNet achieves similar results to LiteDenseNet with 6.06\% parameters and 28.09\% FLOPS. In the small sample categories, compared with other four large networks, the two lightweight networks show good classification properties. Among them, for the 9th category, our proposed network  performs poorly, while HybridSN and LiteDenseNet showed $ 100 \% $ accuracy. It is worth noting that the overall performance of HybridSN is not satisfactory. LeeEtAlNet, the only one that is less computationally expensive than our proposed method, is already unacceptable in terms of accuracy.

\begin{table}[!t]
\centering
\caption{THE CATEGORIZED RESULTS FOR THE \textbf{IP} DATASET USING $ 5 \% $ TRAINING SAMPLES}
\scalebox{0.45}{
\begin{tabular}{c|c|c|c|c|c|c|c|c}
  \toprule
  \toprule
  Class & Color & SVM & ChenEtAlNet & HybridSN & HamidaEtAlNet & LeeEtAlNet & LiteDenseNet & LiteDepthwiseNet \\
  \hline
  1 & \card{myred} & 9.09 & 31.82 & 34.09 & 0.00 & 0.00 & \textbf{100.00} & \textbf{100.00} \\
  2 & \card{mygreen} & 73.03 & 34.83 & 89.17 & 0.00 & 27.93 & 91.97 & \textbf{97.42} \\
  3 & \card{myblue} & 55.00 & 48.81 & 78.83 & 7.10 & 0.00 & \textbf{95.31} & 94.68 \\
  4 & \card{myyellow} & 32.00 & 45.24 & 61.78 & 0.00 & 36.89 & \textbf{98.67} & 98.22 \\
  5 & \card{mypurple} & 89.11 & 46.72 & 79.52 & 0.00 & 0.00 & \textbf{88.89} & \textbf{88.89} \\
  6 & \card{mycyan} & 95.96 & 92.58 & 98.56 & 96.54 & 91.92 & \textbf{99.28} & \textbf{99.28} \\
  7 & \card{mybrown} & 59.26 & 32.00 & 11.11 & 0.00 & 0.00 & \textbf{100.00} & \textbf{100.00} \\
  8 & \card{mygrassgreen} & 99.56 & 86.04 & 96.26 & 92.95 & 95.37 & \textbf{99.78} & \textbf{99.78} \\
  9 & \card{mybluepurple} & 15.79 & 17.65 & \textbf{100.00} & 0.00 & 0.00 & \textbf{100.00} & 86.67 \\
  10 & \card{mycrimson} & 62.19 & 72.46 & 79.52 & 0.00 & 0.00 & \textbf{92.20} & 90.68 \\
  11 & \card{mylightgreen} & 75.17 & 81.49 & 94.85 & 94.60 & 84.18 & 96.12 & \textbf{96.59} \\
  12 & \card{mydarkblue} & 49.02 & 29.54 & 82.77 & 0.00 & 0.00 & 95.74 & \textbf{95.74} \\
  13 & \card{mydarkbrown} & 93.33 & 84.10 & 79.49 & 1.54 & 0.00 & \textbf{96.89} & 95.34 \\
  14 & \card{myblackishgreen} & 93.18 & 78.39 & 98.17 & 99.42 & 96.42 & 96.34 & \textbf{99.58} \\
  15 & \card{mydeeppurple} & 52.59 & 51.39 & 68.94 & 0.27 & 1.91 & \textbf{97.55} & 95.37 \\
  16 & \card{mypink} & 94.32 & 6.90 & 50.00 & 0.00 & 0.00 & 98.85 & \textbf{100.00} \\
  \hline
  OA $ \times 100 $ & & 74.22 & 66.00 & 87.68 & 46.75 & 47.87 & 95.35 & \textbf{96.29} \\
  AA $ \times 100 $ & & 61.68 & 52.52 & 75.19 & 23.08 & 25.57 & \textbf{96.72} & 96.14 \\
  Kappa $ \times 100 $ & & 70.50 & 59.39 & 85.89 & 35.60 & 37.60 & 94.70 & \textbf{95.77} \\
  \hline
  FLOPs & & & 6.030G & 2.483G & 1.148G & \textbf{37.819M} & 1.316G & \textbf{369.331M} \\
  Parameters & & & 1.120M & 8.256M & 6.452M & 5.532M & \textbf{852.309k} & \textbf{51.616k} \\
  \bottomrule
  \bottomrule
\end{tabular}}
\label{table6}
\end{table}

\subsubsection{Classification Maps and Categorized Results for the \textbf{UP} Dataset}

\begin{figure}[!t]
\centering
\subfloat[False color image]{
\includegraphics[width=2.5cm]{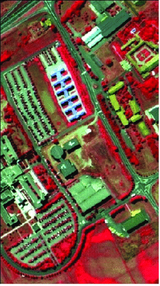}
}
\hfil
\subfloat[GT]{
\includegraphics[width=2.5cm]{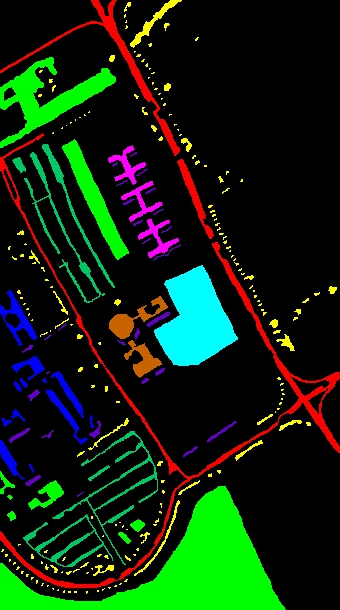}
}
\hfil
\subfloat[SVM]{
\includegraphics[width=2.5cm]{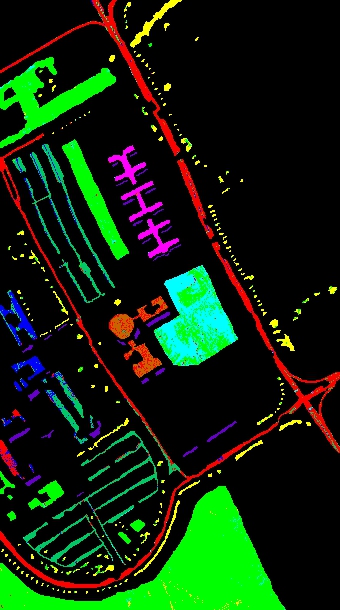}
}
\hfil
\subfloat[ChenEtAlNet]{
\includegraphics[width=2.5cm]{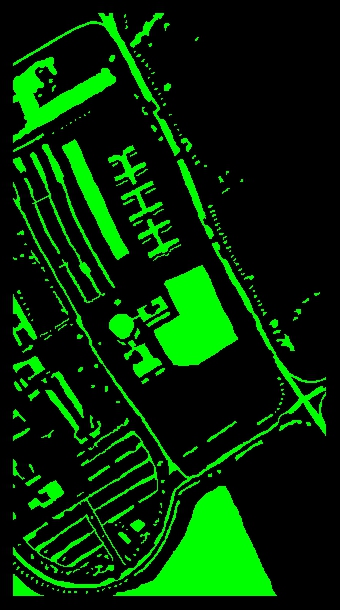}
}
\hfil
\subfloat[HybridSN]{
\includegraphics[width=2.5cm]{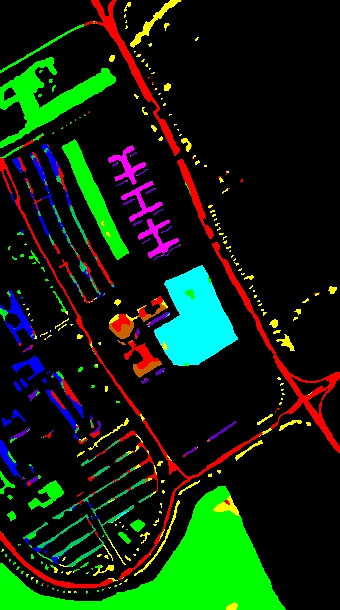}
}
\hfil
\subfloat[HamidaEtAlNet]{
\includegraphics[width=2.5cm]{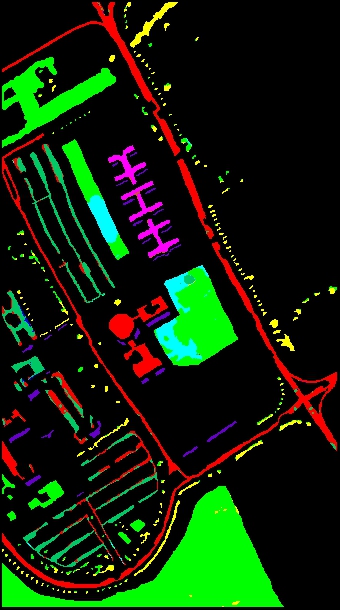}
}\hfil
\subfloat[LeeEtAlNet]{
\includegraphics[width=2.5cm]{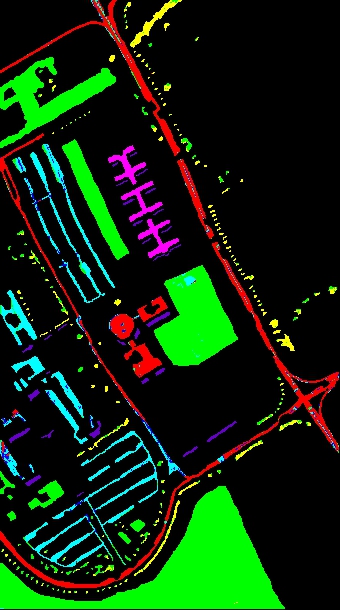}
}
\hfil
\subfloat[LiteDenseNet]{
\includegraphics[width=2.5cm]{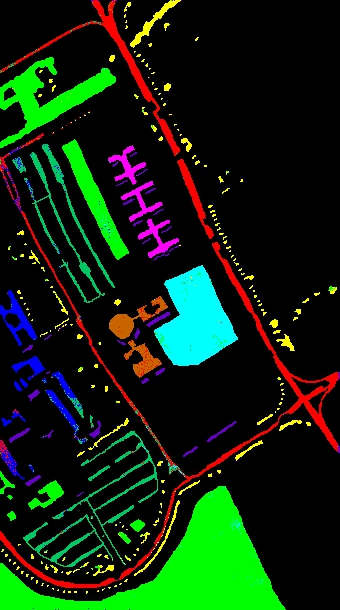}
}
\hfil
\subfloat[LiteDepthwiseNet]{
\includegraphics[width=2.5cm]{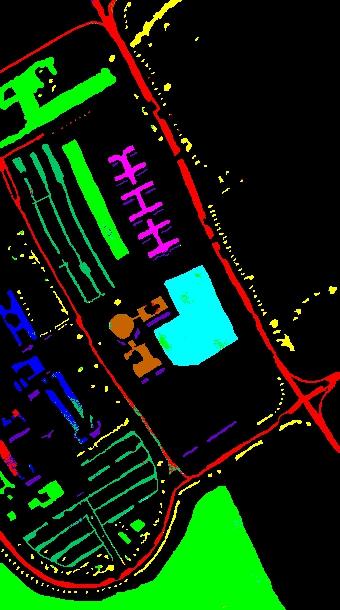}
}
\centering
\caption{Classification maps for the \textbf{UP} dataset using $ 0.5 \% $ of training samples. (a) False-color image. (b) Ground truth (GT). (c)-(i) The classification maps with
disparate algorithms.}
\label{figure12}
\end{figure}

The results of using different algorithms to classify the \textbf{UP} dataset are shown in Table \ref{table7}, and the classification maps of different algorithms and the ground truth are shown in Fig. \ref{figure12}. The main characteristics of this data are the large amount of data and the balanced distribution of categories. In the case of using only 0.5\% of the training samples, our proposed method achieves state-of-the-art performance on OA, AA, and Kappa. It can be clearly seen that some large networks that performed well in the past do not perform well at a small training rate, such as HybridSN, which only achieves an OA of $ 87.68 \% $. Our proposed LiteDepthwiseNet generates the best classification results with the minimal parameters and calculations.

\begin{table}[!t]
\centering
\caption{THE CATEGORIZED RESULTS FOR THE \textbf{UP} DATASET USING $ 0.5 \% $ TRAINING SAMPLES}
\scalebox{0.45}{
\begin{tabular}{c|c|c|c|c|c|c|c|c}
  \toprule
  \toprule
  Class & Color & SVM & ChenEtAlNet & HybridSN & HamidaEtAlNet & LeeEtAlNet & LiteDenseNet & LiteDepthwiseNet \\
  \hline
  1 & \card{myred} & 90.77 & 94.41 & 92.73 & 96.60 & 87.24 & 93.15 & \textbf{97.15} \\
  2 & \card{mygreen} & 98.23 & 68.13 & 98.21 & 90.25 & 98.42 & 99.21 & \textbf{99.77} \\
  3 & \card{myblue} & 51.97 & 0.00 & 78.36 & 2.65 & 7.51 & \textbf{84.84} & 78.10 \\
  4 & \card{myyellow} & 88.20 & \textbf{98.55} & 82.06 & 76.26 & 69.04 & 95.91 & 96.21 \\
  5 & \card{mypurple} & 98.72 & 99.93 & \textbf{100.00} & 96.85 & 99.92 & 99.25 & 99.40 \\
  6 & \card{mycyan} & 62.56 & 0.58 & 97.78 & 25.29 & 2.13 & \textbf{98.35} & 96.77 \\
  7 & \card{mybrown} & 61.12 & 0.00 & 41.50 & 0.23 & 0.00 & 88.84 & \textbf{96.20} \\
  8 & \card{mygrassgreen} & 89.19 & 90.51 & 44.24 & 76.60 & 0.03 & 95.86 & \textbf{97.97} \\
  9 & \card{mybluepurple} & \textbf{100.00} & 95.52 & 73.67 & \textbf{100.00} & 99.89 & 98.51 & 98.72 \\
  \hline
  OA $ \times 100 $ & & 88.11 & 64.50 & 88.28 & 74.75 & 67.35 & 96.60 & \textbf{97.39} \\
  AA $ \times 100 $ & & 73.95 & 54.76 & 78.73 & 56.47 & 46.42 & 94.88 & \textbf{95.59} \\
  Kappa $ \times 100 $ & & 83.90 & 53.90 & 84.27 & 65.40 & 53.50 & 95.50 & \textbf{96.54} \\
  \hline
  FLOPs & & & 1.953G & 1.208G & 591.067M & \textbf{17.88M} & 664.921M & \textbf{187.531M} \\
  Parameters & & & 1.068M & 6.467M & 1.975M & 1.572M & \textbf{437.157k} & \textbf{30.453k} \\
  \bottomrule
  \bottomrule
\end{tabular}}
\label{table7}
\end{table}

\subsubsection{Classification Maps and Categorized Results for the \textbf{PC} Dataset}

\begin{figure}[!t]
\centering
\subfloat[False color image]{
\includegraphics[width=2.5cm]{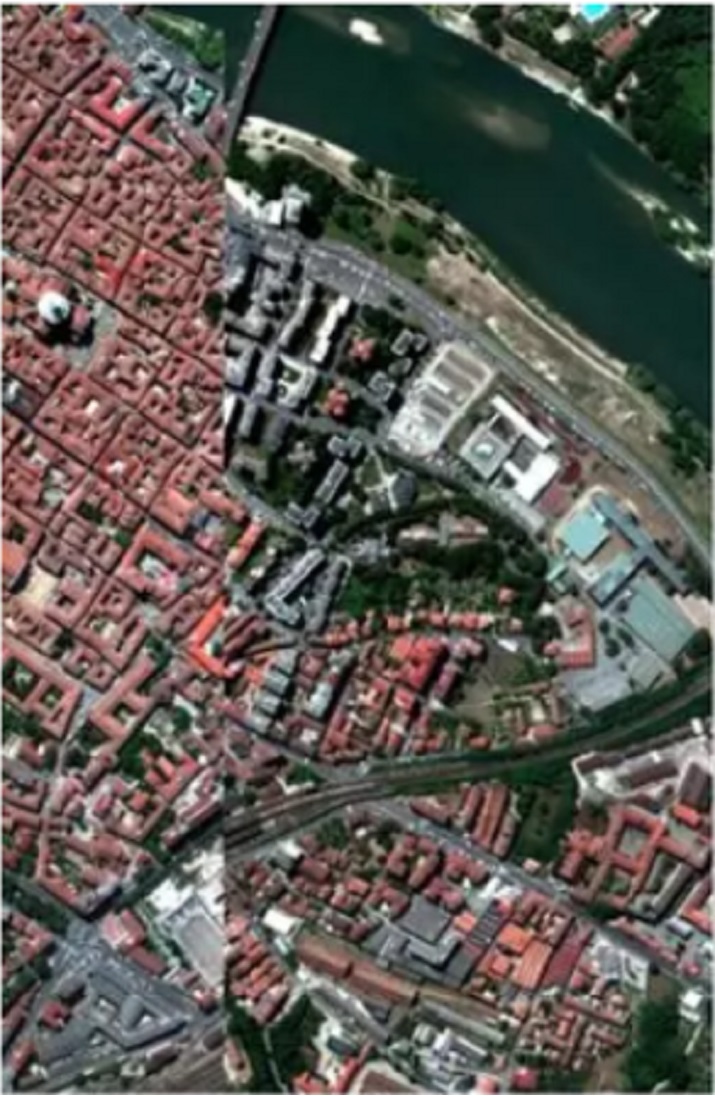}
}
\hfil
\subfloat[GT]{
\includegraphics[width=2.5cm]{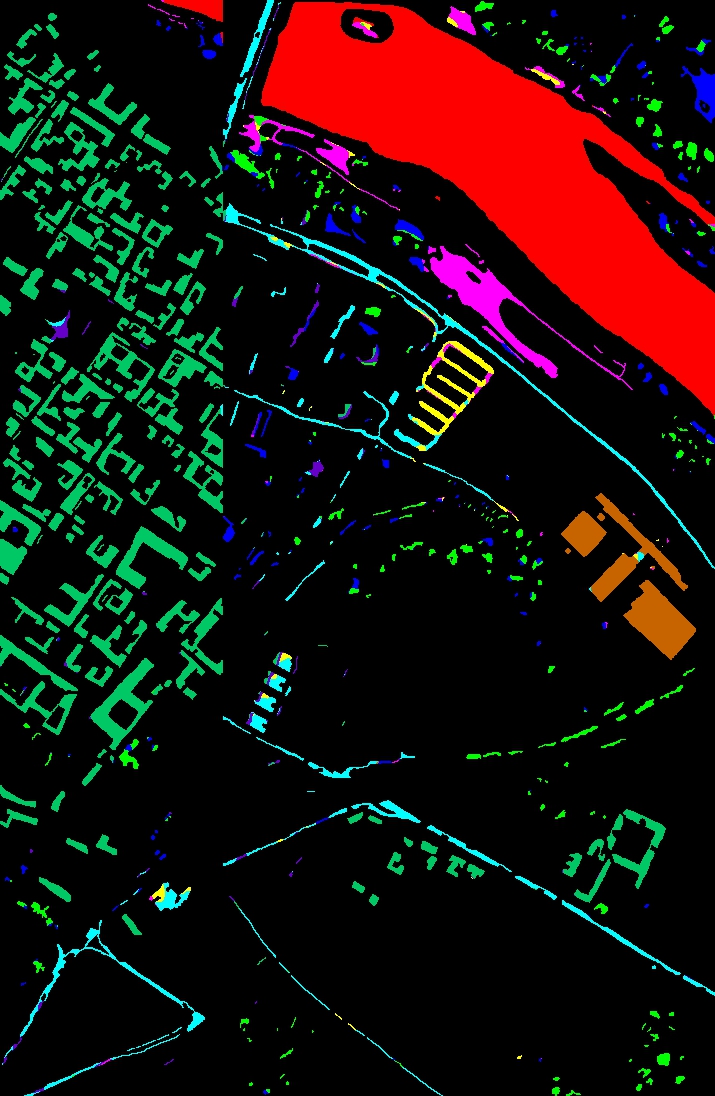}
}
\hfil
\subfloat[SVM]{
\includegraphics[width=2.5cm]{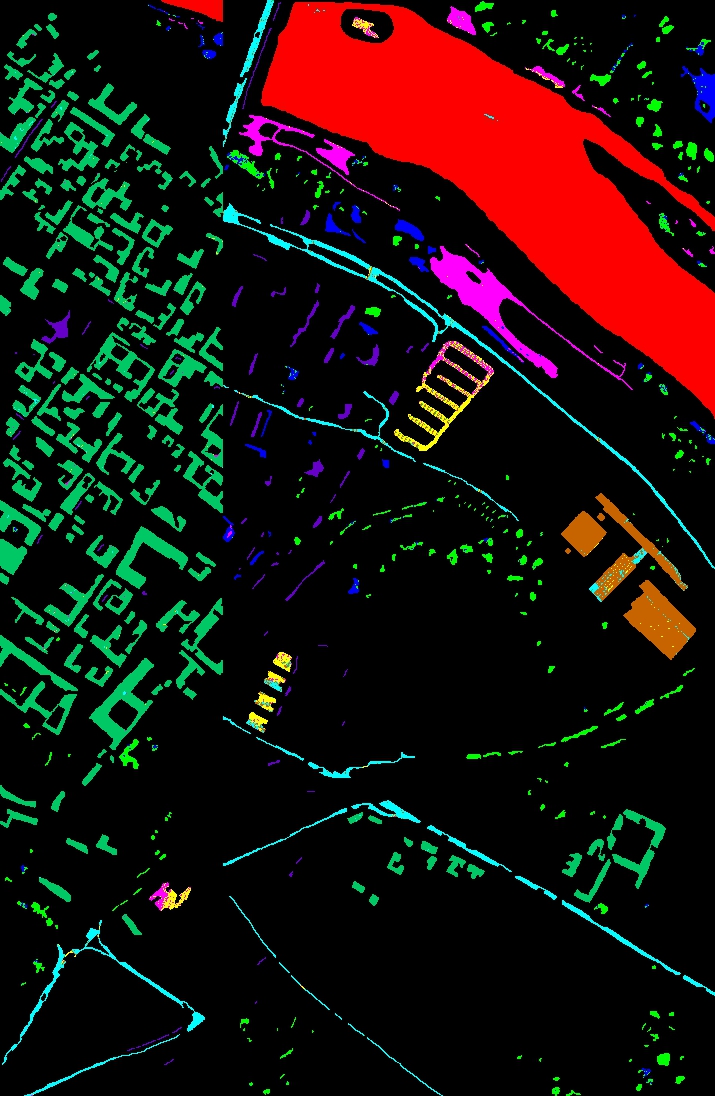}
}
\hfil
\subfloat[ChenEtAlNet]{
\includegraphics[width=2.5cm]{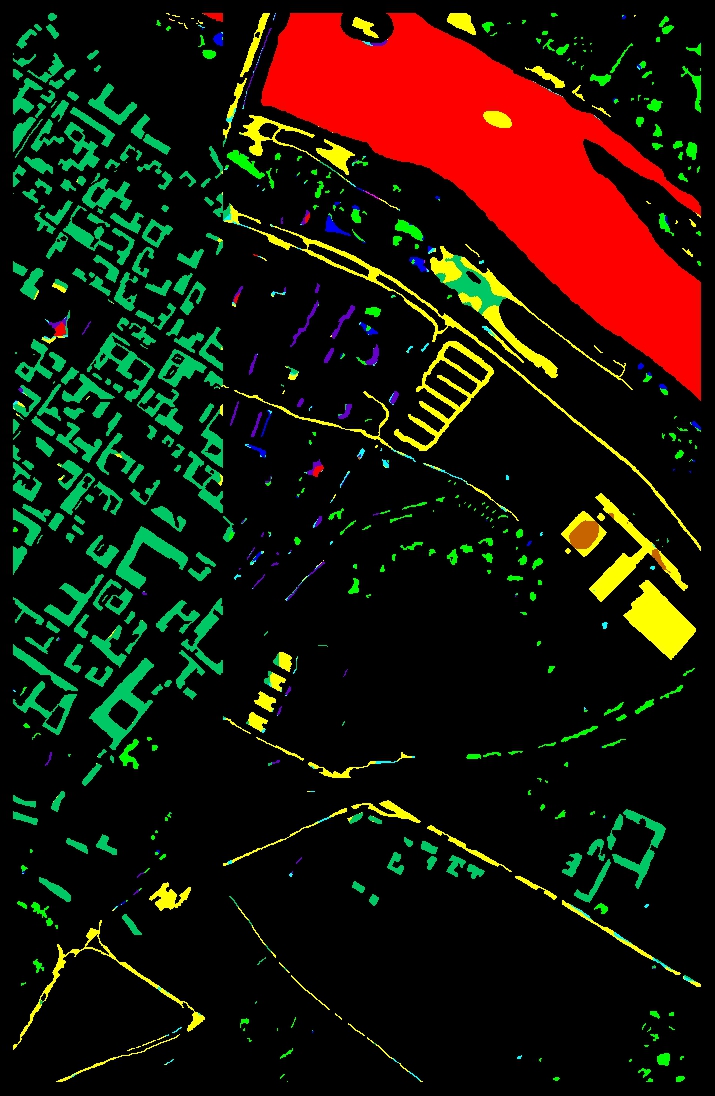}
}
\hfil
\subfloat[HybridSN]{
\includegraphics[width=2.5cm]{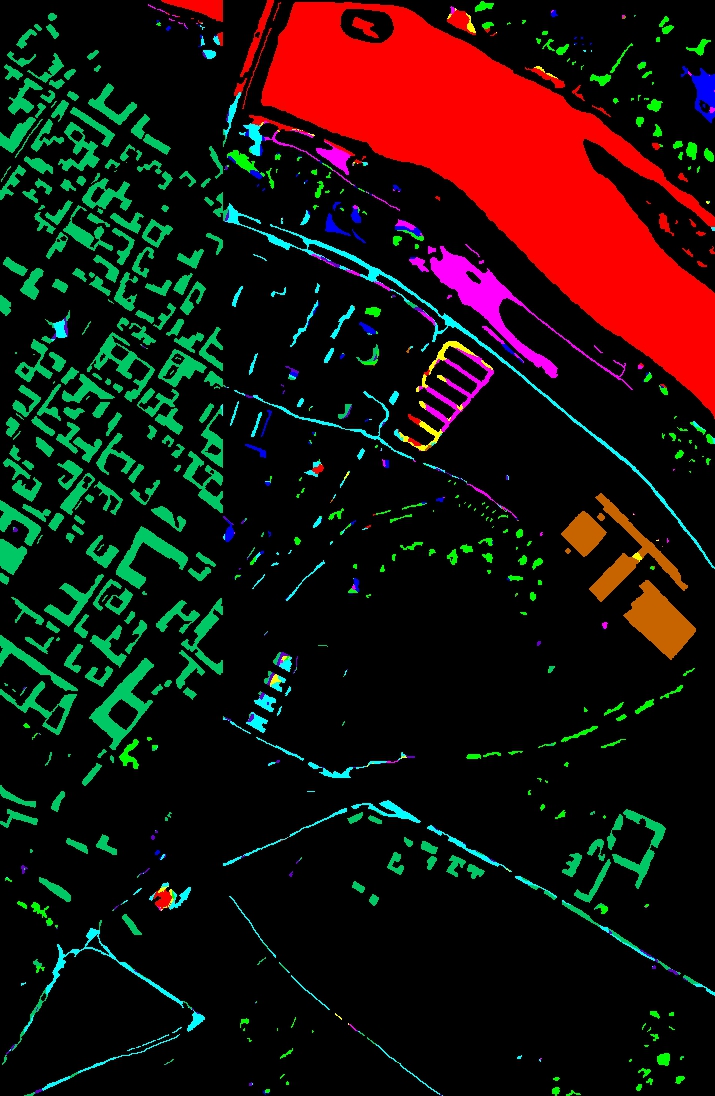}
}
\hfil
\subfloat[HamidaEtAlNet]{
\includegraphics[width=2.5cm]{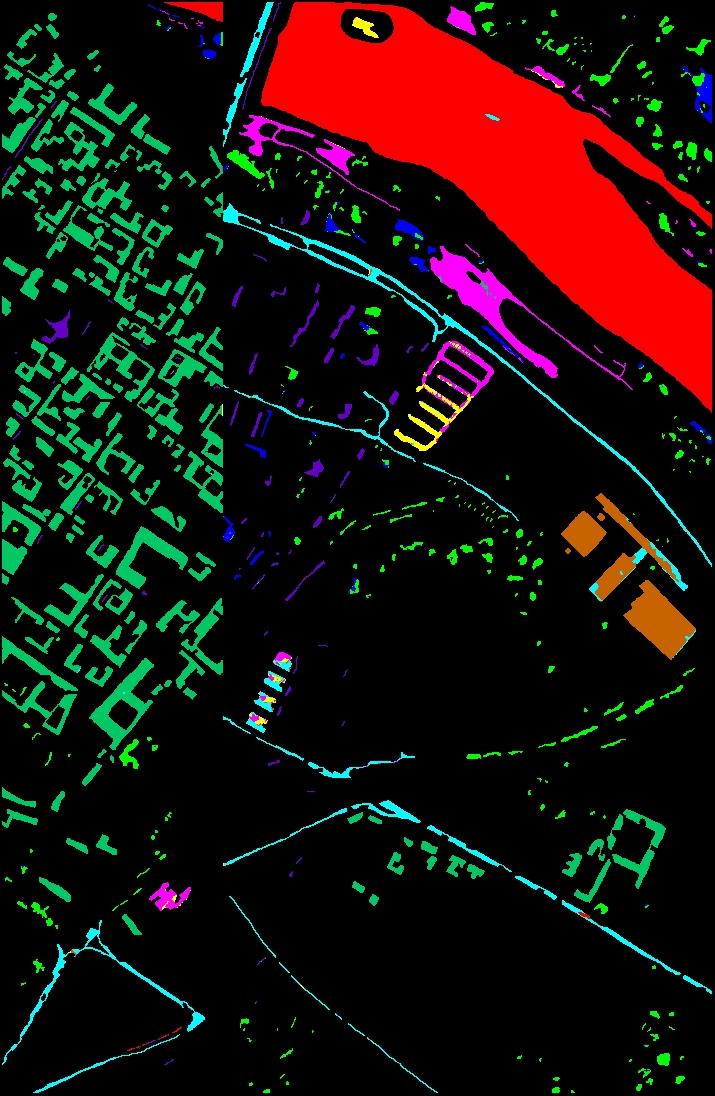}
}
\hfil
\subfloat[LeeEtAlNet]{
\includegraphics[width=2.5cm]{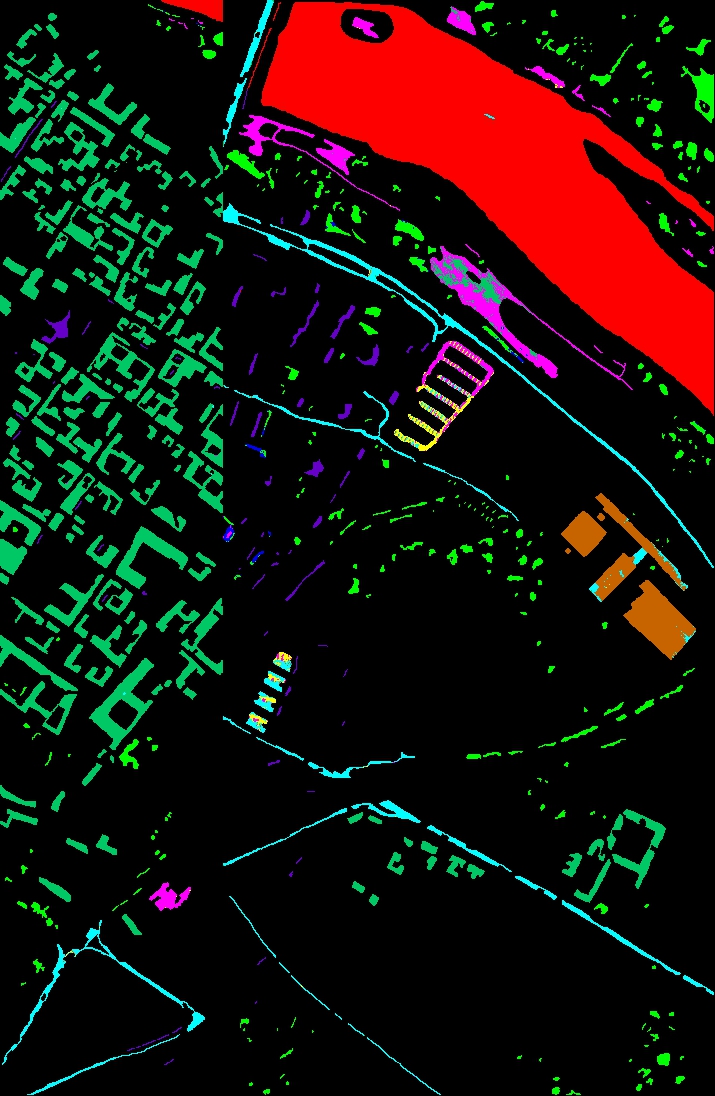}
}
\hfil
\subfloat[LiteDenseNet]{
\includegraphics[width=2.5cm]{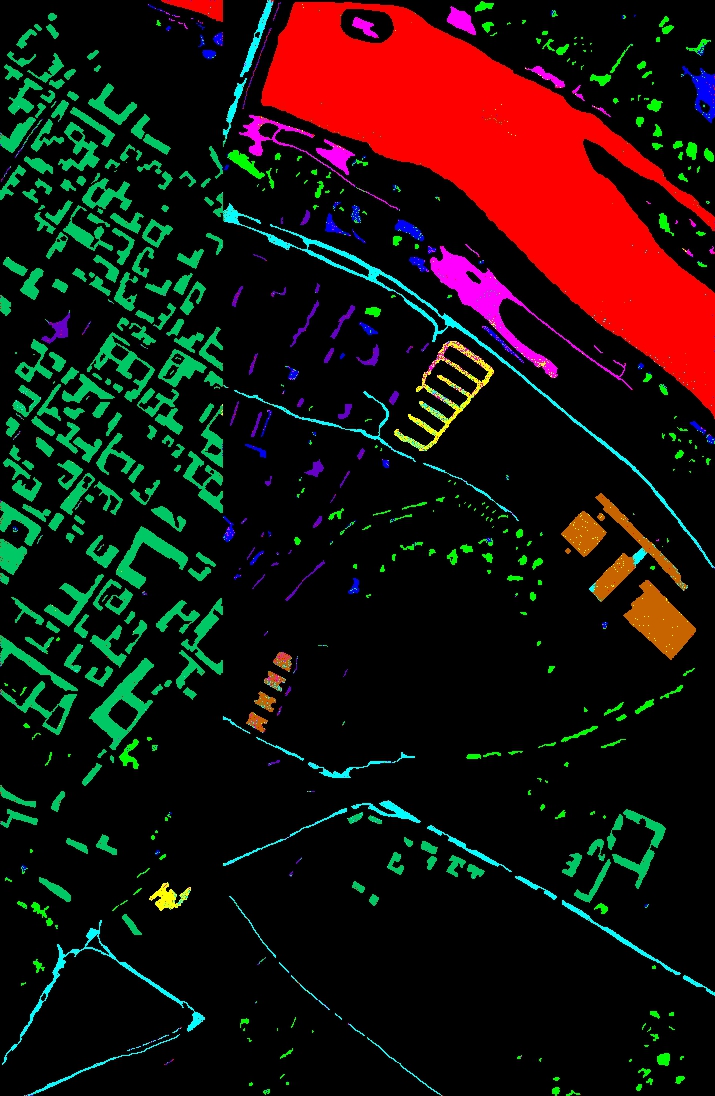}
}
\hfil
\subfloat[LiteDepthwiseNet]{
\includegraphics[width=2.5cm]{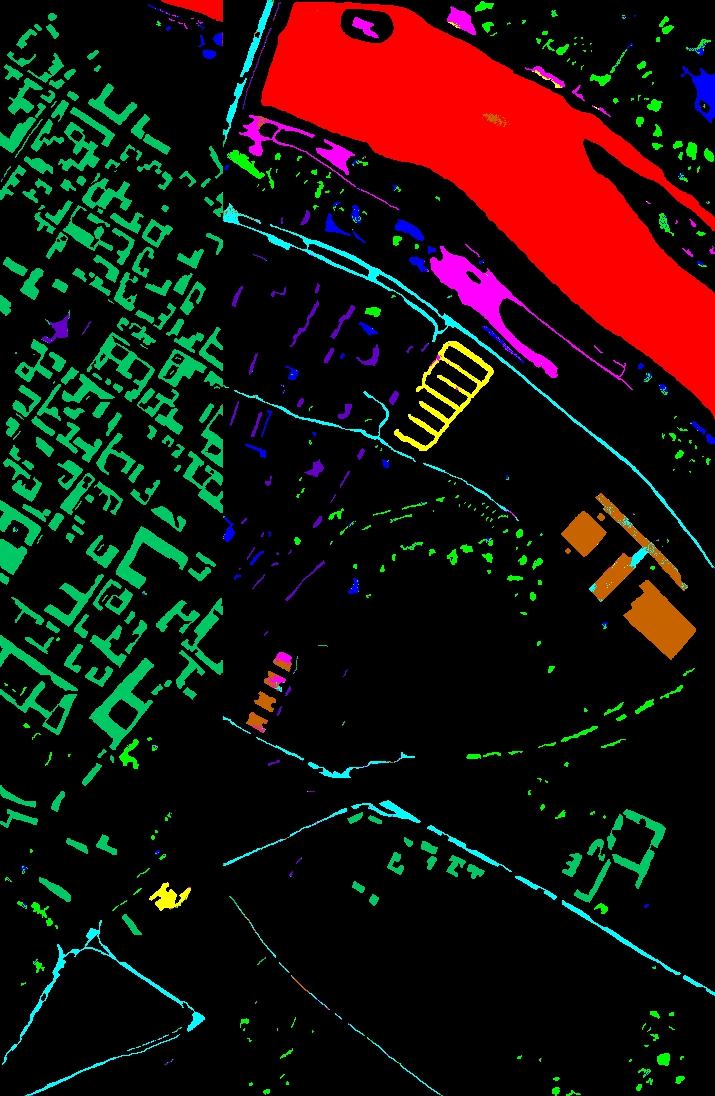}
}
\caption{Classification maps for the \textbf{PC} dataset using $ 0.1 \% $ of training samples. (a) False-color image. (b) Ground truth (GT). (c)-(i) The classification maps with disparate algorithms.}
\label{figure13}
\end{figure}

Table \ref{table8} and Fig. \ref{figure13} show the classification results and maps of different algorithms on the \textbf{PC} dataset, respectively. The overall sample size of the \textbf{PC} dataset is large and basically balanced. Among them, the first and eighth categories are the two categories with the largest number of samples, far exceeding other categories. Therefore, the OA on this dataset is easy to achieve a higher score, but the difficulty lies in the classification of the other seven categories with a small sample size. It can be seen from the above table that, except for the ChenEtAlNet, other six methods achieve good performance and the proposed network achieves the highest OA, AA, Kappa with the smallest amount of parameters, and the amount of calculation is much lower than that of LiteDenceNet. And the LeeEtAlNet still cannot achieve acceptable accuracy on four small sample categories.

\begin{table}[!t]
\centering
\caption{THE CATEGORIZED RESULTS FOR THE \textbf{PC} DATASET USING $ 0.1 \% $ TRAINING SAMPLES}
\scalebox{0.45}{
\begin{tabular}{c|c|c|c|c|c|c|c|c}
  \toprule
  \toprule
  Class & Color & SVM & ChenEtAlNet & HybridSN & HamidaEtAlNet & LeeEtAlNet & LiteDenseNet & LiteDepthwiseNet \\
  \hline
  1 & \card{myred} & 99.96 & 94.73 & \textbf{99.98} & 99.10 & 99.76 & 99.83 & 99.80 \\
  2 & \card{mygreen} & 91.15 & 0.00 & 78.80 & 96.35 & \textbf{99.91} & 94.14 & 91.82 \\
  3 & \card{myblue} & \textbf{93.30} & 0.00 & 59.73 & 67.47 & 10.57 & 84.53 & 93.09 \\
  4 & \card{myyellow} & 59.51 & 0.00 & 31.77 & 39.26 & 32.90 & 71.75 & \textbf{97.50} \\
  5 & \card{mypurple} & 96.03 & 0.00 & 70.66 & 94.34 & 83.13 & 95.80 & \textbf{97.52} \\
  6 & \card{mycyan} & 96.92 & 0.00 & 75.13 & 97.91 & \textbf{99.03} & 97.27 & 96.61 \\
  7 & \card{mybrown} & 83.35 & 0.00 & 85.98 & 79.59 & 81.42 & \textbf{91.97} & 90.87 \\
  8 & \card{mygrassgreen} & 99.38 & 0.00 & 99.96 & 99.22 & 99.95 & 99.37 & \textbf{99.98} \\
  9 & \card{mybluepurple} & \textbf{99.93} & 98.35 & 6.54 & 97.79 & 96.28 & 86.24 & 85.33 \\
  \hline
  OA & & 97.28 & 44.47 & 91.46 & 95.98 & 94.99 & 97.59 & \textbf{98.24} \\
  AA & & 81.95 & 9.47 & 67.62 & 77.10 & 70.29 & 91.21 & \textbf{94.73} \\
  Kappa $ \times 100 $ & & 96.20 & 0.00 & 87.70 & 94.30 & 92.90 & 96.58 & \textbf{97.51} \\
  \hline
  FLOPs & & & 1.953G & 1.95G & 591.067M & \textbf{17.880M} & 651.353M & \textbf{183.743M} \\
  Parameters & & & 1.068M & 6.448M & 1.975M & 1.572M & \textbf{428.517k} & \textbf{30.021k} \\
  \bottomrule
  \bottomrule
\end{tabular}}
\label{table8}
\end{table}

\subsection{Investigation of the Effects of Different Loss Functions}

Table \ref{table9} gives the experimental results of the proposed algorithm with different loss functions on the three dataset. It can be seen that the focal loss has a significant improvement on the performance of hyperspectral classification. Specifically, for the \textbf{IP} dataset, due to the imbalance of sample categories, the balanced cross entropy loss increases by about $ 14 \% $ AA compared with the standard cross entropy loss, which benefits from increasing the loss weight of small class samples. Based on this, focal loss increases the weight of samples that are difficult to classify, and further improves OA, AA, and Kappa, increasing by $ 2.63 \% $, $ 0.68 \% $, and $ 3 \% $ respectively. Compared with the growth of balanced cross-entropy loss, the performance improvement of focal loss obviously benefits from difficult samples. For the \textbf{UP} and \textbf{PC} datasets, there is no problem of sample imbalance, so standard cross entropy and balanced cross entropy show similar performance, while focal loss still has a significant improvement. In the \textbf{UP} dataset, compared with the CEL and the BCEL, the OA, AA, Kappa of FL increase by about $ 2.77 \% , 10.22 \% $, and $ 3.72 \% $, respectively. The overall classification of the \textbf{PC} dataset is less difficult and the room for improvement is small, but there is still $0.37\%, 2.7\%$, and $0.54\%$ increase on the three quantitative indicators respectively. In summary, the mining of difficult samples by focal loss improves the performance to a certain extent for HSI classification.

\begin{table}[!t]
\centering
\caption{THE OA COMPARISON OF LITEDEPTHWISENET BETWEEN DIFFERENT LOSS FUNCTIONS}
\scalebox{0.65}{
\begin{tabular}{c|c|c|c|c|c|c|c|c|c}
  \toprule
  \toprule
  \multirow{2}*{Dataset} & \multicolumn{3}{c|}{Standard Cross Entropy} &  \multicolumn{3}{c|}{Balanced Cross Entropy} & \multicolumn{3}{c}{Focal Loss} \\
  \cline{2-10}
   & OA & AA & Kappa & OA & AA & Kappa & OA & AA & Kappa \\
  \hline
  IP & 92.13 & 81.56 & 91.03 & 93.66 & 95.46 & 92.77 & 96.29 & 96.14 & 95.77 \\
  UP & 94.63 & 85.38 & 92.83 & 94.63 & 85.40 & 92.83 & 97.39 & 95.59 & 96.54 \\
  PC & 97.92 & 92.02 & 97.04 & 97.91 & 91.99 & 97.03 & 98.24 & 94.73 & 97.51 \\
  \bottomrule
  \bottomrule
\end{tabular}}
\label{table9}
\end{table}

\subsection{Investigation of the $ \gamma $ of Focal Loss}

\begin{figure}[H]
  \centering
  \includegraphics[scale=0.5]{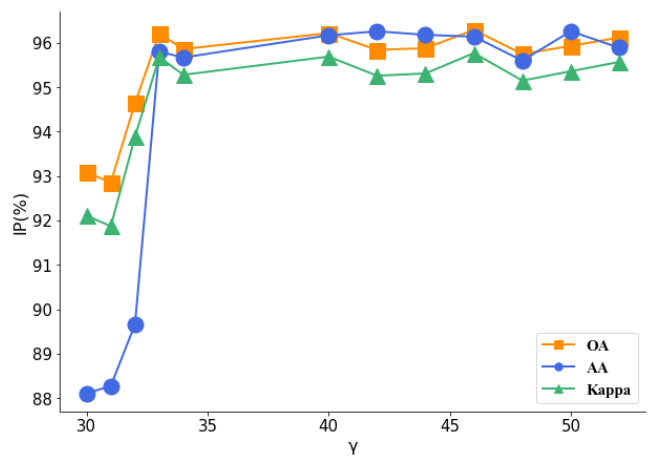}
  \caption{The OA, AA, and Kappa of LiteDepthwiseNet with different values of $ \gamma $ on the \textbf{IP} dataset.}
  \label{figure14}
\end{figure}

\begin{figure}[H]
  \centering
  \includegraphics[scale=0.5]{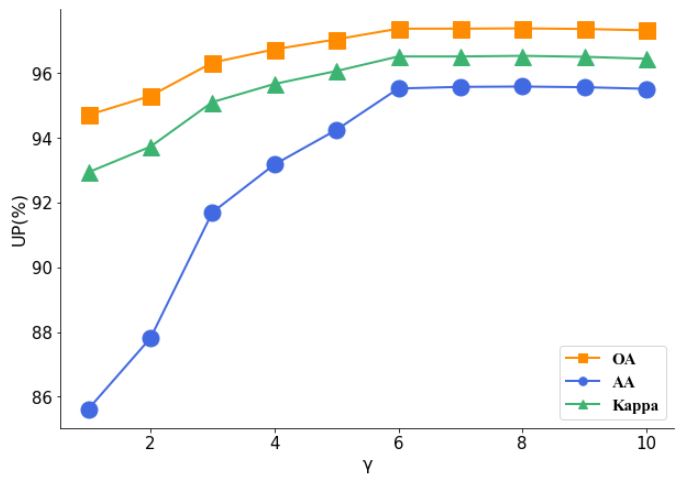}
  \caption{The OA, AA, and Kappa of LiteDepthwiseNet with different values of $ \gamma $ on the \textbf{UP} dataset.}
  \label{figure15}
\end{figure}

\begin{figure}[H]
  \centering
  \includegraphics[scale=0.5]{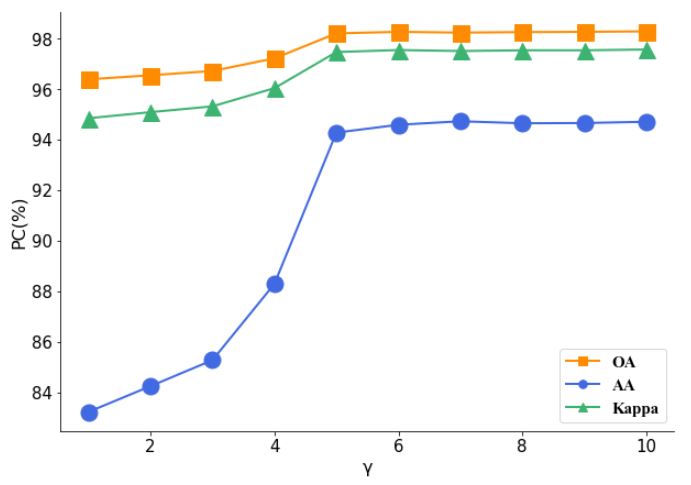}
  \caption{The OA, AA, and Kappa of LiteDepthwiseNet with different values of $ \gamma $ on the \textbf{PC} dataset.}
  \label{figure16}
\end{figure}

Fig. \ref{figure14}, Fig. \ref{figure15} and Fig. \ref{figure16} show the effect of different values of $ \gamma $ in the proposed LiteDepthwiseNet for OA, AA, and Kappa on the three datasets. For different datasets, the optimal weight and degree of influence are different. Of these, the value of $ \gamma $ has a particularly significant impact on AA because AA is particularly sensitive to the accuracy of small sample categories (often difficult-classified samples), and adjusting $ \gamma $ can significantly enhance attention to difficult-classified samples. In our experiment, the optimal values of $ \gamma $ for the three datasets \textbf{IP}, \textbf{UP}, and \textbf{PC} are 46, 8, and 7, respectively.

\section{Conclusion}\label{V}

In this paper, we propose an HSI classification network, LiteDepthwiseNet, which can achieve state-of-the-art performance with minimal parameters and computational cost. The network is constructed by the modified 3D depthwise convolution, which greatly reduces the model parameters and computational overhead. At the same time, this modification enhances the linearity of the model so our model avoids the phenomenon of overfitting on datasets with very small samples. In addition, compared with the CE loss, we find that the focal loss can improve the limited performance of the model. As a lightweight network, the proposed LiteDepthwiseNet can be deployed on devices with low computational power for HSI classification. We believe that further work can be devoted to improving the generalization ability of the model. When the distribution of the training and verification set is different, it should also achieve good performance. When the untrained data meets the category that has been seen, it should also show excellent performance, so it can be deployed in industry to improve productivity.

\section*{Acknowledgments}
The authors would like to thank Prof. D. Landgrebe for providing the Indian Pines data set,
and Prof. P. Gamba for providing the University of Pavia and Pavia Center data sets.

\bibliographystyle{IEEEtran}
\bibliography{mybibfile}

\begin{thebibliography}{10}
\providecommand{\url}[1]{#1}
\csname url@samestyle\endcsname
\providecommand{\newblock}{\relax}
\providecommand{\bibinfo}[2]{#2}
\providecommand{\BIBentrySTDinterwordspacing}{\spaceskip=0pt\relax}
\providecommand{\BIBentryALTinterwordstretchfactor}{4}
\providecommand{\BIBentryALTinterwordspacing}{\spaceskip=\fontdimen2\font plus
\BIBentryALTinterwordstretchfactor\fontdimen3\font minus
  \fontdimen4\font\relax}
\providecommand{\BIBforeignlanguage}[2]{{%
\expandafter\ifx\csname l@#1\endcsname\relax
\typeout{** WARNING: IEEEtran.bst: No hyphenation pattern has been}%
\typeout{** loaded for the language `#1'. Using the pattern for}%
\typeout{** the default language instead.}%
\else
\language=\csname l@#1\endcsname
\fi
#2}}
\providecommand{\BIBdecl}{\relax}
\BIBdecl

\bibitem{Li2019}
Z.~Li, L.~Huang, and J.~He, ``A multiscale deep middle-level feature fusion
  network for hyperspectral classification,'' \emph{Remote Sensing}, vol.~11,
  no.~6, p. 695, Mar. 2019.

\bibitem{Yokoya2016}
N.~Yokoya, J.~Chan, and K.~Segl, ``Potential of resolution-enhanced
  hyperspectral data for mineral mapping using simulated enmap and sentinel-2
  images,'' \emph{Remote Sensing}, vol.~8, no.~3, p. 172, Feb. 2016.

\bibitem{Yang2017}
X.~{Yang} and Y.~{Yu}, ``Estimating soil salinity under various moisture
  conditions: An experimental study,'' \emph{IEEE Transactions on Geoscience
  and Remote Sensing}, vol.~55, no.~5, pp. 2525--2533, May. 2017.

\bibitem{Xie2019}
W.~{Xie}, T.~{Jiang}, Y.~{Li}, X.~{Jia}, and J.~{Lei}, ``Structure tensor and
  guided filtering-based algorithm for hyperspectral anomaly detection,''
  \emph{IEEE Transactions on Geoscience and Remote Sensing}, vol.~57, no.~7,
  pp. 4218--4230, Jul. 2019.

\bibitem{Li2019Overview}
S.~{Li}, W.~{Song}, L.~{Fang}, Y.~{Chen}, P.~{Ghamisi}, and J.~A.
  {Benediktsson}, ``Deep learning for hyperspectral image classification: An
  overview,'' \emph{IEEE Transactions on Geoscience and Remote Sensing},
  vol.~57, no.~9, pp. 6690--6709, Sep. 2019.

\bibitem{Qian2001}
Q.~Du and C.~I. Chang, ``A linear constrained distance-based discriminant
  analysis for hyperspectral image classification,'' \emph{Pattern
  Recognition}, vol.~34, no.~2, pp. 361--373, Jun. 2001.

\bibitem{Melgani2004}
F.~{Melgani} and L.~{Bruzzone}, ``Classification of hyperspectral remote
  sensing images with support vector machines,'' \emph{IEEE Transactions on
  Geoscience and Remote Sensing}, vol.~42, no.~8, pp. 1778--1790, Aug. 2004.

\bibitem{Li2010}
J.~{Li}, J.~M. {Bioucas-Dias}, and A.~{Plaza}, ``Semisupervised hyperspectral
  image segmentation using multinomial logistic regression with active
  learning,'' \emph{IEEE Transactions on Geoscience and Remote Sensing},
  vol.~48, no.~11, pp. 4085--4098, Nov. 2010.

\bibitem{Zhong2012}
Y.~{Zhong} and L.~{Zhang}, ``An adaptive artificial immune network for
  supervised classification of multi-/hyperspectral remote sensing imagery,''
  \emph{IEEE Transactions on Geoscience and Remote Sensing}, vol.~50, no.~3,
  pp. 894--909, Mar. 2012.

\bibitem{Peng2015}
J.~{Peng}, Y.~{Zhou}, and C.~L.~P. {Chen}, ``Region-kernel-based support vector
  machines for hyperspectral image classification,'' \emph{IEEE Transactions on
  Geoscience and Remote Sensing}, vol.~53, no.~9, pp. 4810--4824, Sep. 2015.

\bibitem{Benediktsson2005}
J.~A. {Benediktsson}, J.~A. {Palmason}, and J.~R. {Sveinsson}, ``Classification
  of hyperspectral data from urban areas based on extended morphological
  profiles,'' \emph{IEEE Transactions on Geoscience and Remote Sensing},
  vol.~43, no.~3, pp. 480--491, Mar. 2005.

\bibitem{Li2013}
J.~{Li}, P.~R. {Marpu}, A.~{Plaza}, J.~M. {Bioucas-Dias}, and J.~A.
  {Benediktsson}, ``Generalized composite kernel framework for hyperspectral
  image classification,'' \emph{IEEE Transactions on Geoscience and Remote
  Sensing}, vol.~51, no.~9, pp. 4816--4829, Sep. 2013.

\bibitem{Kang2014}
X.~{Kang}, S.~{Li}, and J.~A. {Benediktsson}, ``Spectral-spatial hyperspectral
  image classification with edge-preserving filtering,'' \emph{IEEE
  Transactions on Geoscience and Remote Sensing}, vol.~52, no.~5, pp.
  2666--2677, May. 2014.

\bibitem{Chen2011}
Y.~{Chen}, N.~M. {Nasrabadi}, and T.~D. {Tran}, ``Hyperspectral image
  classification using dictionary-based sparse representation,'' \emph{IEEE
  Transactions on Geoscience and Remote Sensing}, vol.~49, no.~10, pp.
  3973--3985, Oct. 2011.

\bibitem{Fang2014}
L.~{Fang}, S.~{Li}, X.~{Kang}, and J.~A. {Benediktsson}, ``Spectral-spatial
  hyperspectral image classification via multiscale adaptive sparse
  representation,'' \emph{IEEE Transactions on Geoscience and Remote Sensing},
  vol.~52, no.~12, pp. 7738--7749, Dec. 2014.

\bibitem{Fang2017}
L.~{Fang}, C.~{Wang}, S.~{Li}, and J.~A. {Benediktsson}, ``Hyperspectral image
  classification via multiple-feature-based adaptive sparse representation,''
  \emph{IEEE Transactions on Instrumentation and Measurement}, vol.~66, no.~7,
  pp. 1646--1657, Jul. 2017.

\bibitem{Peng2019}
J.~Peng, X.~Jiang, N.~Chen, and H.~Fu, ``Local adaptive joint sparse
  representation for hyperspectral image classification,''
  \emph{Neurocomputing}, vol. 334, no. MAR.21, pp. 239--248, Jan. 2019.

\bibitem{Peng2017}
J.~{Peng}, H.~{Chen}, Y.~{Zhou}, and L.~{Li}, ``Ideal regularized composite
  kernel for hyperspectral image classification,'' \emph{IEEE Journal of
  Selected Topics in Applied Earth Observations and Remote Sensing}, vol.~10,
  no.~4, pp. 1563--1574, Apr. 2017.

\bibitem{Chen2014}
Y.~{Chen}, Z.~{Lin}, X.~{Zhao}, G.~{Wang}, and Y.~{Gu}, ``Deep learning-based
  classification of hyperspectral data,'' \emph{IEEE Journal of Selected Topics
  in Applied Earth Observations and Remote Sensing}, vol.~7, no.~6, pp.
  2094--2107, Jun. 2014.

\bibitem{Zhang2017}
X.~{Zhang}, Y.~{Liang}, C.~{Li}, N.~{Huyan}, L.~{Jiao}, and H.~{Zhou},
  ``Recursive autoencoders-based unsupervised feature learning for
  hyperspectral image classification,'' \emph{IEEE Geoscience and Remote
  Sensing Letters}, vol.~14, no.~11, pp. 1928--1932, Nov. 2017.

\bibitem{Pan2017}
B.~{Pan}, Z.~{Shi}, and X.~{Xu}, ``R-vcanet: A new deep-learning-based
  hyperspectral image classification method,'' \emph{IEEE Journal of Selected
  Topics in Applied Earth Observations and Remote Sensing}, vol.~10, no.~5, pp.
  1975--1986, May. 2017.

\bibitem{Cao2018}
X.~{Cao}, F.~{Zhou}, L.~{Xu}, D.~{Meng}, Z.~{Xu}, and J.~{Paisley},
  ``Hyperspectral image classification with markov random fields and a
  convolutional neural network,'' \emph{IEEE Transactions on Image Processing},
  vol.~27, no.~5, pp. 2354--2367, May. 2018.

\bibitem{lee2016}
H.~{Lee} and H.~{Kwon}, ``Contextual deep cnn based hyperspectral
  classification,'' in \emph{2016 IEEE International Geoscience and Remote
  Sensing Symposium (IGARSS)}, Nov. 2016, pp. 3322--3325.

\bibitem{chen2016}
Y.~{Chen}, H.~{Jiang}, C.~{Li}, X.~{Jia}, and P.~{Ghamisi}, ``Deep feature
  extraction and classification of hyperspectral images based on convolutional
  neural networks,'' \emph{IEEE Transactions on Geoscience and Remote Sensing},
  vol.~54, no.~10, pp. 6232--6251, Oct. 2016.

\bibitem{Benhamida2018}
A.~{Ben Hamida}, A.~{Benoit}, P.~{Lambert}, and C.~{Ben Amar}, ``3-{D} deep
  learning approach for remote sensing image classification,'' \emph{IEEE
  Transactions on Geoscience and Remote Sensing}, vol.~56, no.~8, pp.
  4420--4434, Aug. 2018.

\bibitem{Roy2020}
S.~K. {Roy}, G.~{Krishna}, S.~R. {Dubey}, and B.~B. {Chaudhuri},
  ``Hybrid{S}{N}: Exploring 3{D}--2{D} cnn feature hierarchy for hyperspectral
  image classification,'' \emph{IEEE Geoscience and Remote Sensing Letters},
  vol.~17, no.~2, pp. 277--281, Feb. 2020.

\bibitem{Li2020}
R.~Li and C.~Duan, ``Litedensenet: A lightweight network for hyperspectral
  image classification,'' \emph{ArXiv}, vol. abs/2004.08112, Apr. 2020.

\bibitem{Howard2017MobileNets}
A.~G. Howard, M.~Zhu, B.~Chen, D.~Kalenichenko, W.~Wang, T.~Weyand,
  M.~Andreetto, and H.~Adam, ``Mobilenets: Efficient convolutional neural
  networks for mobile vision applications,'' \emph{ArXiv}, vol. abs/1704.04861,
  Apr. 2017.

\bibitem{Pelee2018}
R.~J. Wang, X.~Li, and C.~X. Ling, ``Pelee: A real-time object detection system
  on mobile devices,'' in \emph{Advances in Neural Information Processing
  Systems 31}, 2018, pp. 1963--1972.

\bibitem{Li2020Classification}
R.~Li, S.~Zheng, C.~Duan, Y.~Yang, and X.~Wang, ``Classification of
  hyperspectral image based on double-branch dual-attention mechanism
  network,'' \emph{Remote Sensing}, vol.~12, no.~3, p. 582, Feb. 2020.

\bibitem{Ye20193D}
R.~Ye, F.~Liu, and L.~Zhang, ``3{D} depthwise convolution: Reducing model
  parameters in 3{D} vision tasks,'' in \emph{Canadian Conference on Artificial
  Intelligence}, 2019, pp. 186--199.

\bibitem{Lin2017Focal}
T.~Y. Lin, P.~Goyal, R.~Girshick, K.~He, and P.~Dollár, ``Focal loss for dense
  object detection,'' \emph{IEEE Transactions on Pattern Analysis \& Machine
  Intelligence}, vol.~42, no.~2, pp. 318--327, 2020.

\end{thebibliography}

\end{document}